\documentclass[12pt]{article}
\usepackage{amsmath,slashed, amssymb,enumerate}

\def\){\right)} 
\def\({\left(} 
\def\tr{\mbox{tr}}
\def\str{\mbox{str}}
\def\sdet{\mbox{sdet}}
\def\Dslash{{\rm D}\!\!\!\!/\,}

\newcommand{\eqn}[1]{\label{eq:#1}}
\newcommand{\refeq}[1]{(\ref{eq:#1})} 
\newcommand{\eq}{eq.~\refeq}
\newcommand{\eqs}[2]{eqs.~(\ref{eq:#1}-\ref{eq:#2})}

\newcommand{\Eq}{Eq.~\refeq} 

\newcommand{\beq}{\begin{eqnarray}}
\newcommand{\eeq}{\end{eqnarray}}
\newcommand{\mcal}[1]{{\mathcal #1}}

\begin{document}

\title{ \vspace{0cm}
Partially quenched chiral perturbation theory\\
without $\Phi_0$\\
\vspace{.5cm}
}
\author{Stephen Sharpe\thanks{sharpe@phys.washington.edu}
\,\, and Noam Shoresh\thanks{shoresh@phys.washington.edu} \\ \\
 Physics Department, University of Washington \\
        Seattle, WA 98195-1560}
\date{August 2, 2001}
\maketitle
\vspace{.5cm}
\setcounter{page}{0}
\thispagestyle{empty}

\begin{abstract}
This paper completes the argument that lattice simulations  of
partially quenched QCD can provide quantitative information about QCD itself,
with the aid of partially quenched chiral perturbation theory. 
A barrier to doing this has been the inclusion of $\Phi_0$,
the partially quenched generalization of the $\eta'$, 
in previous calculations in the partially quenched effective theory.
This invalidates the low energy perturbative expansion,
gives rise to many new unknown parameters, and makes it impossible to
reliably calculate the relation between the 
partially quenched theory and low energy QCD. 
We show that it is straightforward and natural to formulate
partially quenched chiral perturbation theory without $\Phi_0$,
and that the resulting theory contains the effective theory for QCD
without the $\eta'$. We also show that previous results, obtained
including $\Phi_0$, can be reinterpreted as applying to the theory
without $\Phi_0$.
We contrast the situation with that in the quenched effective theory,
where we explain why it is necessary to include $\Phi_0$.
 We also compare the derivation of chiral perturbation
theory in partially quenched QCD with the standard derivation in unquenched
QCD. We find that the former cannot be justified as rigorously as the latter,
because of the absence of a physical
Hilbert space. Finally, we present an encouraging
result: unphysical double poles in certain correlation functions in partially
quenched chiral perturbation theory can be shown to be a property
of the underlying theory, given 
only the symmetries and some plausible assumptions.
\end{abstract}
\vspace{0.25in}
\begin{flushright}
UW/PT 01-17 
\end{flushright}

\newpage

\begin{section}{Introduction}

Numerical simulations of lattice QCD are hampered by the
difficulty of including loops of light quarks.
This has forced the use of approximations to the
fermion determinant: the quenched approximation (setting the determinant
to a constant), and, more recently, the partially quenched (PQ) approximation
(including the determinant but with sea quark masses different from,
and typically larger than, those of the valence quarks).
While all such simulations correspond to unphysical theories,
they are not all equally unphysical.
It has been argued recently that PQ
simulations can be used to obtain 
physical parameters if the quarks are light enough that
one can use chiral perturbation theory to describe the low energy
properties of the theory~\cite{shshI,ckn,shshII}. 
The only approximation necessary is the truncation
of chiral perturbation theory. On the other hand, if the sea-quarks are
too heavy, then partial quenching is an uncontrolled approximation
whose results will at best be a qualitative guide to those
in the physical theory.

The main purpose of this paper is to complete the theoretical argument
justifying the use of PQ simulations to obtain physical parameters.
The missing ingredient in the arguments presented in Refs.~\cite{shshI,shshII}
concerns the flavor-singlet field, $\Phi_0$, 
which is the generalization of the $\eta'$ in PQ theories.
This field must be kept in the low-energy effective theory for
quenched QCD, because its correlation functions contain poles
at the masses of the light pseudo-Goldstone mesons.
The same holds true for the PQ theory with heavy sea quarks.
The need to include $\Phi_0$ invalidates the standard
chiral power counting and introduces additional coupling
constants in the chiral Lagrangian. Consequently, progress with this 
theory can only be made by making further assumptions. 

On the other hand, in QCD the $\eta'$ is not a pseudo-Goldstone boson,
the corresponding field need not be included in the chiral
Lagrangian, and chiral perturbation theory can be developed as 
a low energy expansion~\cite{weinberg,gassleutannphy,gassleut}.
What we demonstrate here
is that the situation is similar for PQ QCD 
with sea (and valence) quarks in the
chiral regime: It can be described by a chiral Lagrangian in which
$\Phi_0$ is absent, and for which standard power-counting applies. 
As stressed in Refs.~\cite{shshI,shshII},
an important corollary is
that the parameters of the PQ chiral Lagrangian
are the same as those in the chiral Lagrangian for QCD.
This is what is needed to show that PQ simulations can be
used to extract physical parameters.

Another result shown here is of a more technical nature.
All previous calculations using PQ chiral perturbation theory
have included the $\Phi_0$ field, and thus suffer from the
problems described above.
Here we show that these problems can be avoided by sending
the $\Phi_0$ mass parameter, $m_0$, to infinity,
because  this is mathematically equivalent to 
considering the theory in which $\Phi_0$ is absent.
In this way, the results of the previous calculations from PQ chiral
perturbation theory can be reinterpreted as 
applying to the theory for which the matching to QCD is immediate.
This observation also justifies the ad-hoc prescriptions for 
integrating out $\Phi_0$ that were used 
previously~\cite{shZ,shPQ,GLPQ,shshII}.

A secondary purpose of this paper is to discuss the theoretical foundations of
chiral perturbation theory for PQ theories, 
building on the work of \cite{BGPQ,verbaarschot}.  We recall the line of
reasoning used to construct the chiral Lagrangian for QCD,
examine the extent to which it applies also to PQ QCD,
and point out the gaps that make PQ chiral
perturbation theory stand on less secure grounds.  
We are able to show, however, that a
signature prediction of PQ chiral effective theories, namely the presence of
double pole contributions to flavor singlet correlators, can be derived in the
underlying theory from the assumption that valence flavor 
non-singlet correlators have
single poles. The latter assumption is well tested by numerical simulations.

This paper is organized as follows.  In the next section we review the
definition of PQ theories, and discuss their symmetries. 
The set of two-point functions containing light poles is identified in
Section~\ref{sec:lowpoles} for both PQ and quenched theories.
In Section~\ref{sec:effL} we construct the PQ effective Lagrangian
that reproduces the previously determined pole structure, and discuss
its theoretical foundations.
In Section~\ref{sec:phi0} we explain how chiral perturbation theory
including $\Phi_0$ is equivalent to that excluding $\Phi_0$
if one sends $m_0\to\infty$.
We then discuss the pole structure of flavor-singlet correlators in
Section~\ref{sec:poles}, and summarize our conclusions in 
Section~\ref{sec:conclusions}.
 
Five appendices contain technical details.  \ref{app:fakesym} describes the
true symmetries of the PQ QCD partition function, derives the resulting Ward
identities, and shows how these are in fact the same as those obtained
assuming a ``fake'' symmetry group.  \ref{app:defs} collects useful results on
the diagonal generators of graded Lie groups.  \ref{app:alter} gives an
alternative argument for why one does not need to include $\Phi_0$ in PQ chiral
perturbation theory.  \ref{app:Ginverse} concerns a useful result about the
limit $m_0\to\infty$.  Finally, \ref{app:Gstructure} derives the constraints on
neutral pion correlators using graded symmetries.

\end{section}

\begin{section}{Partially quenched theories and their symmetries}
\label{sec:PQ}

We consider a theory with $N$ sea quarks and $N_V$ valence quarks,
which is viewed as a tool to study an unquenched theory
with $N$ quarks. We will shortly consider all quarks, sea and
valence, to be light, although this is not needed to set up
the definition of the theory.
Clearly, the theories we have in mind are those with 
$N=2$ (treating only the up and down quarks as light)
and $N=3$ (also including the strange quark---whose status as
a light quark is less clear).
The number of valence quarks one usually uses is $N_V=2$
(needed to discuss simple meson properties),
or $N_V=3$ (needed for baryon properties).

In practice, one obtains PQ results by generating gauge configurations
including in the weight a determinant representing the effect
of $N$ sea quarks, and then calculating quark propagators on these
background gauge fields using masses which are different from those 
of the sea quarks.
This can be represented theoretically by Morel's construction involving
bosonic spin-1/2 ghost fields~\cite{Morel}, 
in which the Euclidean partition function is
\beq
Z&=&\int D[\psi\bar{\psi}\tilde{\psi}\bar{\tilde{\psi}}A] 
\exp \( -S_{G}-\int \left[\bar{\psi}\(\Dslash+m\)\psi+
\bar{\tilde{\psi}}\(\Dslash+\tilde{m}\)\tilde{\psi}\) \right]
       \nonumber      \\    &=&\int
D[A]\exp\(-S_G\)\frac{\det\(\Dslash+m\)}{\det\(\Dslash+\tilde{m}\)}
\eqn{Z} \,.
\eeq 
Here $S_G$ is the gauge action, $\psi$ an $N_V+N$-dimensional
column vector containing the quark fields, with mass matrix $m$,
and $\tilde{\psi}$ an $N_V$-dimensional vector of ghost quarks,
with mass matrix $\tilde{m}$.
For each valence quark there is a ghost quark with the same mass,
so that the valence quark determinant cancels precisely against 
that from the ghost quarks. The complete mass matrix is
\beq
M=\( \begin{array}{cc} 
    m & 0 \\ 
    0 & \tilde{m} 
    \end{array} \)
 =\mbox{diag}\(
m_{V1},\ldots,m_{VN_V};m_{S1},\ldots,m_{SN};
{m}_{V1},\ldots,{m}_{VN_V}\)
\,.
\eqn{M} 
\eeq 
In the following, we will take all non-zero entries of $M$ to be
real and positive.

Our discussion will concern the theory obtained in the
continuum limit of PQ simulations. 
This allows us to take $\Dslash$ as having standard continuum properties:
it is an anti-hermitian operator which connects left-handed fields
to right-handed fields and {\em vice-versa}.
In this way we avoid the issue of how best to discretize fermions,
and of possible difficulties in simulating odd numbers of dynamical quarks.
We simply assume that these difficulties have been overcome,
and that the lattice simulations are done close enough to the continuum
limit that eq.~(\ref{eq:Z}) represents them
up to small corrections, suppressed by powers of the lattice spacing,
which can be extrapolated away.%
\footnote{Some comments on the impact
of  $O(a)$ corrections are made in Ref.~\cite{shshII}.}
Indeed by using overlap fermions, or other fermions with an exact chiral
symmetry, one can presumably formulate the
discussion of symmetries at non-zero lattice spacing.

Correlation functions of quark and ghost fields are defined, as usual,
by introducing source terms into $Z$. 
Note that if one considers correlation functions involving only 
sea-quark fields,
one obtains exactly the result of the unquenched theory
with $N$ sea quarks, because the valence and ghost determinants 
cancel~\cite{BGPQ}.
Furthermore, if one of the valence quarks has the same mass as a
sea quark, then it can replace that sea quark in correlation functions
without changing the result~\cite{BGPQ}. 

To discuss symmetries,
it is useful to collect quarks and ghost quarks into a single 
$N+2 N_V$-dimensional vector, $Q$, defined by
\beq
Q^T = (\psi^T, \tilde\psi^T) \,.
\eeq
The fermionic part of the action then takes the standard form
\beq
\eqn{SF}
S_F &=& \int \overline{Q} (\Dslash + M) Q  \\
&=& \int\left( 
\overline{Q}_L \Dslash Q_L + \overline{Q}_R \Dslash Q_R
+\overline{Q}_R M Q_L + \overline{Q}_L M Q_R \right)
\,, \nonumber
\eeq
where the projections are defined by
\beq
Q_{L,R} = \frac{(1\pm\gamma_5)}{2} Q \,,\quad
\overline{Q}_{L,R} = \overline{Q}\frac{(1\mp\gamma_5)}{2}  \,.
\eeq

The symmetry group of the action \eq{SF}, when $M\to0$, appears to be
the graded group
\beq
U(N_V+N|N_V)_L \otimes U(N_V+N|N_V)_R \,,
\eqn{Usym}
\eeq
under which the fields transform in the usual way
\beq
Q_{L,R}(x) \longrightarrow U_{L,R} Q_{L,R}(x) \,,\qquad
\overline{Q}_{L,R}(x) \longrightarrow \overline{Q}_{L,R}(x) U_{L,R}^\dagger
 \,,
\eqn{Utrans}
\eeq
with $U_{L,R} \in U(N_V+N|N_V)$.
As noted in Ref.~\cite{verbaarschot}, however, this is not the 
correct symmetry
because the functional integral over the bosonic spin-1/2 fields
converges only if they are properly constrained 
(and if the mass matrix is positive definite).
As a result of the constraint, right- and left-handed fields cannot be rotated
independently in the usual way. 
Nevertheless, it turns out that one can proceed as if the
symmetry group were (\ref{eq:Usym}), as long as one only considers small
transformations. This is explained in \ref{app:fakesym}.
In particular we show that one obtains the
correct vector and axial Ward identities
if one pretends that the action has the symmetry group \eq{Usym},
rather than the actual symmetry group.
In what follows we mostly use the ``fake'' symmetries, since this emphasizes
the similarities to the development for unquenched QCD, and is usually
simpler. We use the real symmetries only when global aspects of the symmetry
group are important.

Certain of the transformations in \eq{Utrans} are anomalous,
since they do not leave the measure invariant.
Removing these requires that we impose the constraint
\beq
\sdet\ U_L = \sdet\ U_R \,,
\eqn{sdetcond}
\eeq
where ``sdet'' is the invariant ``super-determinant''.\footnote{%
Here the super-determinent is defined by
$\sdet\ U=\exp \str \ln U$, with the supertrace being
\begin{equation}
\str \left( {\begin{matrix}
A&B\\
C&D\\
\end{matrix}} \right)
  = \tr A - \tr D \,,
\end{equation} 
the blocks corresponding to the 
$N_V+N$ quark and $N_V$ ghost coordinates, respectively.}
The overall result is that the chiral symmetry group of the 
massless PQ theory can be taken to be the group proposed in \cite{BGPQ}
\beq
SU(N_V+N|N_V)_L \otimes SU(N_V+N|N_V)_R \otimes U(1)_V \,.
\eqn{pretendsymm}
\eeq
Here the $U(1)_V$ transformations are common overall phase rotations of the
right- and left-handed fields.\footnote{We are ignoring
the fact that globally $U(N_V+N|N_V)=[SU(N_V+N|N_V)\otimes U(1)]/Z_N$,
i.e. it is a coset rather than a direct product. This is irrelevant
for small transformations.}

The form \eq{pretendsymm} reflects the fact that
for the PQ case (i.e. $N>0$), the $U(1)$ factor can be chosen
to be a flavor singlet, and thus to 
commute with all elements of $SU(N_V+N|N_V)$.
In particular, the anomalous symmetry current 
can be chosen to be the flavor singlet, 
$j_{A\mu}^{(I)}  = \overline Q \gamma_\mu  \gamma_5 Q$
(using the notation of \ref{app:defs}, \eq{AxialCurrent}).
This has the same quantum numbers as the pseudoscalar density
$\overline Q \gamma _5 Q$,
which is the interpolating field associated with $\Phi_0$~\cite{BGPQ}---the 
focus of much of the subsequent discussion.

The symmetry structure is different for the fully quenched theory, $N=0$.
In particular, as noted in Ref.~\cite{BGlat92},
the flavor singlet $U(1)$ is part of $SU(N_V|N_V)$,
while the $U(1)$ with non-unit superdeterminant cannot
be chosen to be a flavor singlet, and so does not commute with
all elements of $SU(N_V|N_V)$. This is explained more fully
in \ref{app:defs}, where we collect some results on the generators
of graded groups.
The net effect is that the $SU(N_V|N_V)$ and $U(1)$ factors form,
locally, a semi-direct product, and
the chiral symmetry group can be taken to be
\begin{align}
\left[\operatorname{SU} (N_V |N_V )_L  \otimes 
\operatorname{SU} (N_V |N_V )_R\right] \ltimes  U(1)_V. 
\eqn{pretendsymmQ}
\end{align}
As discussed below, the difference between this group and \eq{pretendsymm}
is the mathematical result which underlies the need to include
the $\Phi_0$ field in the quenched,
but not the PQ, chiral Lagrangian.

\end{section}

\begin{section}{Symmetry breaking and the need for $\Phi_0$}
\label{sec:lowpoles}

As could already be seen in the previous section, the introduction of valence
and ghost quarks modifies the standard flavor symmetry structure of the
theory, and consequently the formulation of the chiral effective theory.  One
striking example is the non-decoupling of $\Phi_0$ in the fully quenched
theory. The main goal of this section is to analyze this phenomenon in some
detail, and to demonstrate why it does not carry over to PQ theories. To study
this, we investigate which two-point correlation functions contain poles at low
energies. In other words, we find the degrees of freedom that one must include
in a low energy effective theory for quenched and PQ QCD. 

Throughout this section we consider the PQ theory
in the chiral limit. This allows
us to use the condensate as an order parameter, and to use Goldstone's
theorem to determine which channels have massless poles.
In the usual way, the chiral limit is to be approached by working
at non-zero quark masses and then taking the masses to zero after
the volume has been sent to infinity. In the PQ theory there is, however,
a subtlety concerning the chiral limit. As noted in Ref.~\cite{shPQ},
chiral perturbation theory predicts that
the PQ theory is singular if one sends valence quark masses to zero
with fixed non-zero sea-quark masses.
In particular, the condensate itself is singular in this limit.
This singularity is similar to that which occurs in the quenched 
theory~\cite{BGQ,BGlat92,SSlat92}.
To avoid this singularity, one should send all the quark masses
to zero simultaneously with fixed ratios.\footnote{%
This limiting proceedure also avoids divergences from exact zero modes
in topologically non-trivial sectors.}

Such a limiting proceedure is not available for the quenched theory.
Thus, in the following, when
we refer to the quenched theory, we have in mind
working close to, but not in, the chiral limit. Since our focus here
is on the PQ theory, we do not revisit the subtleties associated with
the quenched chiral limit.

\begin{subsection}{Symmetry breaking pattern}
\label{subsec:orderparam}
As in QCD, we choose the vacuum expectation value (VEV) 
\begin{align}
\Omega _{ab}  \equiv \left\langle {\overline Q _a Q_b } \right\rangle
\end{align}
as an order parameter for chiral spontaneous symmetry breaking.

We first consider the vector symmetries. It was shown long ago by 
Vafa and Witten 
that vector symmetries do not spontaneously break in vector-like
gauge theories\cite{VafaWitten}. Their derivation does not make use of Hilbert
space states and operators, and relies only on the fact that the quark
determinant leads to a real and positive measure in the functional integral
over the gauge fields. This is still true for quenched and PQ QCD, and the
Vafa-Witten
result still holds. Consequently, $\Omega_{ab}$ is invariant under vector
transformations. It is in fact easier to see what this implies for
the related quantity
\begin{align}
\widetilde\Omega _{ab}  
\equiv \left\langle {Q_{b\tau } \overline Q _{a\tau } } \right\rangle 
\eqn{VEV} 
\end{align}
($\tau$ is a Dirac-color index that must still be contracted to form a
Euclidean scalar). This transforms
under (``fake'') vector transformations in the following way:
\begin{align}
\widetilde\Omega  \to V\widetilde\Omega V^\dag  ,\quad V \in 
SU\left( {N_V  + N|N_V } \right).\eqn{VEVtrans}
\end{align}
The invariance of $\Omega$ under \eq{VEVtrans} leads to
\begin{align}
\widetilde\Omega  = \omega \delta _{ab}
\eqn{OmegatildeVEV}
\end{align}
where $\omega$ is a constant.  
Interchanging $Q$ and $\overline Q$ fields, we obtain
\begin{align}
\Omega _{ab}  =  - \omega \delta _{ab} \varepsilon _a \eqn{PQVEV}
\end{align}
where we introduce the notation
\beq
\varepsilon _a  = \begin{cases}
  1& a\text{ is a quark index},\\
   - 1& a\text{ is a ghost index}.\\ 
\end{cases}
\eeq

The question of whether the axial symmetry is spontaneously broken or not
depends on the value of $\omega$. The sea sector of PQ theories is equivalent
to unquenched QCD with $N$ quark flavors. An important implication of this
fact is that the spontaneous breakdown of axial symmetries in QCD,
signaled by the non-vanishing of $\left\langle {\bar qq} \right\rangle $, is
duplicated in the sea sector of PQ theories, 
thus implying that $\omega\ne 0$.  In other words, given that there
is spontaneous chiral symmetry
breaking in QCD, the symmetry breaking pattern of PQ QCD is known:\footnote{%
Here we are assuming $N\ge 2$, since there is no chiral symmetry to break
for $N=1$ QCD.}
\begin{align}
SU(N_V  + N|N_V )_L  \otimes SU(N_V  + N|N_V )_R  
\otimes U(1)_V  \to SU(N_V  + N|N_V )_V  \otimes U(1)_V .\eqn{SBPQ}
\end{align}

This argument does not carry over to the quenched theories, because there is
no QCD-like sea sector. For quenched theories the spontaneous breaking of
axial symmetries is an additional assumption---though one that is supported by
numerical evidence. If we make that assumption, the symmetry breaking pattern
for quenched QCD is
\begin{align}
\left[SU(N_V |N_V )_L  \otimes SU(N_V |N_V )_R  \right] \ltimes U(1)_V  
\to SU(N_V |N_V )_V  \ltimes U(1)_V .\eqn{SBQ}
\end{align}

\end{subsection}

\begin{subsection}{Low energy poles from symmetries}
\label{sec:lowpolesfromsymm}

Once the symmetry breaking pattern has been established, it is a standard
result in field theory that the number of pseudo-Goldstone bosons is
determined by the number of non-anomalous generators that act non-trivially on
the vacuum. For the case at hand one thus expects $(2N_V+N)^2-1$ Goldstone
particles for PQ theory. Though the group structure in the quenched case is
slightly different, the counting argument still gives
$(2N_V)^2-1$, which is the same expression as for PQ theories with $N=0$.
There is, however, a significant difference between the two cases. As the
following more careful analysis shows, the simple counting argument correctly
predicts the number of Goldstone particles {\em only} in the PQ case. For the
quenched case it turns out that there are $(2N_V)^2$ fields which exhibit long
range correlations. The additional field that is needed for quenched QCD is
none other than $\Phi_0$.

\smallskip
Consider the two-point correlation function between an axial current
$j_{A\mu}^{(T)}(x)$ (defined in \eq{AxialCurrent}) 
and a pseudoscalar density
$\phi^{(T')}(0)  = \overline Q \gamma _5 T'Q(0)$,
\begin{align}
C_\mu ^{(T,T')} (x) = 
\left\langle {j_{A\mu }^{(T)}(x) \phi ^{(T')} (0)} \right\rangle .
\end{align}
Here $T$ and $T'$ label generators of the symmetry group. From Euclidean
invariance, its Fourier transform must have the form
\begin{align}
\tilde C_\mu ^{(T,T')} (p) = i p_\mu  F^{(T,T')} (p^2 ),\eqn{Cform}
\end{align}
where $F$ is an unknown function.  Next, consider the Ward identity which
follows from applying an infinitesimal axial transformation with generator $T$
to $\left\langle {\phi ^{(T')} (0)} \right\rangle $:
\begin{align}
\partial _\mu C_\mu ^{(T,T')} (x) = - \delta (x)\left\langle {\delta _A^{(T)}
\phi ^{(T')} (0)} \right\rangle.\eqn{onlyward}
\end{align}
(for the definition of $\delta_A^{(T)} \phi$ see \eq{deltaOA}.)
Together, \eqs{Cform}{onlyward} give
\begin{align}
p^2 F^{(T,T')} (p^2 ) =
  - \left\langle {\delta _A^{(T)} \phi ^{(T')} (0)} \right\rangle. 
\eqn{ward}
\end{align}
A straightforward calculation of the right hand side, with the use of the VEV
(\ref{eq:PQVEV}), gives
\begin{align}
p^2 F^{(T,T')} (p^2 ) =  2\omega\operatorname{str} (TT') \eqn{ward2}.
\end{align}
(Note that the order of $T$ and $T'$ is important for graded generators.)
\Eq{ward2} relies on the fact that $T$ generates a true (non-anomalous)
symmetry of the theory. For quenched and PQ theories the non-anomalous
symmetries satisfy $\operatorname{str} (T) = 0$ . Consequently,
$\tilde C_\mu ^{(T,T')} (p)$
has a pole at $p=0$ for any $T$ and $T'$ for which 
$\operatorname{str}(T)=0$ and $\operatorname{str}(TT')\ne0$.

It is straightforward to show that, for both quenched and PQ theories, all the
flavor off-diagonal (``charged'')
generators give rise to light poles, like in QCD, and that they
do not mix with the neutral ones. Subtleties only arise in the neutral sector,
and to study them we use a specific choice of diagonal generators that is
presented in \ref{app:defs}. There we find that, for the PQ theory,
\begin{align}
{\operatorname{str} (T_a T_b )} = \operatorname{diag} (\underbrace {1, \ldots
,1}_{N_V + N - 1},\underbrace { - 1, \ldots , - 1}_{N_V - 1}, -
1,1).\tag{\ref{eq:PQstr}}
\end{align}
Here the $T_a$ are diagonal members of the algebra, all straceless but the
last one, $T_{2N_V+N}\propto I$, which generates the anomalous
$\mbox{U}(1)_A$.  Applying \eq{PQstr} to \eq{ward2} shows that a complete set
of (flavor neutral) fields that give rise to long range correlations in
two-point functions is
\begin{align}
\overline Q \gamma _5 T_a Q,\quad a = 1, \ldots ,2N_V  + N - 1.
\end{align}
Note that for $N_V$ of these fields, the coefficient of the massless pole
has an unphysical sign, corresponding to the entries
with $-1$ in \eq{PQstr}.
The symmetries do not imply that $\phi^{(T_{2N_V+N})}=\phi^{(I)}/\sqrt{N}$,
has a pole at $p=0$. Consequently, to describe the long distance parts of the
two-point functions in the PQ theory we do not need to consider correlators
involving the corresponding field, $\Phi_0$.

In the quenched case \eq{PQstr} is replaced by:
\begin{align}
{\operatorname{str} (T_a T_b )}  = \left( {\begin{matrix}
1&{}&{}&{}&{}&{}&{}&{}\\
{}& \ddots &{}&{}&{}&{}&{}&{}\\
{}&{}&1&{}&{}&{}&{}&{}\\
{}&{}&{}&{ - 1}&{}&{}&{}&{}\\
{}&{}&{}&{}& \ddots &{}&{}&{}\\
{}&{}&{}&{}&{}&{ - 1}&{}&{}\\
{}&{}&{}&{}&{}&{}&0&1\\
{}&{}&{}&{}&{}&{}&1&0\\
\end{matrix}} \right). \tag{\ref{eq:Qstr}}
\end{align}

Again, the generator of the anomalous $\mbox{U}(1)_A$ is the last one,
$T_{2N_V}\propto \bar I$ (defined in \ref{app:defs}). All the other generators
are straceless, including $T_{2N_V-1}=I$.  One obtains a new non-trivial
equation from \eq{ward2} by choosing $T=I$, which is non-anomalous (a fact
unique to quenched QCD), and $T'=\bar I$. We thus find that the Fourier
transform of
\begin{align}
\left\langle {j_{A\mu }^{(I)} (x)\phi ^{(\bar I)} (0)} \right\rangle 
\eqn{offdiagWI}
\end{align}
has a pole at $p=0$. Since $\Phi_0$ has the same quantum numbers as 
\begin{align}
j_{A\mu }^{(I)}  = \overline Q \gamma _\mu  \gamma _5 Q
\end{align}
we conclude that it must be included in the effective Lagrangian in order to
correctly reproduce the low energy behavior of quenched QCD.

Let us summarize why one needs to include $\Phi_0$ in quenched,
but not PQ, chiral perturbation theory.
A field should be included if either the corresponding pseudoscalar
density or axial current can be shown to couple to a Goldstone boson.
In unquenched and PQ theories, the argument given above shows that
the coupling of a density with flavor generator $T$ implies the
coupling of the corresponding current [since $\str(T_a T_b)$ is diagonal].
The peculiarity of the quenched theory is that the current with
the same quantum numbers as $\Phi_0$, 
i.e. the flavor singlet  $j_{A\mu}^{(I)}$, 
is {\em not anomalous} and thus should couple to a Goldstone boson.
This is despite the fact that the corresponding density, 
$\overline Q\gamma_5 Q$, is invariant under non-anomalous transformations.
Conversely, the anomalous current,
$j_{A\mu}^{(\bar I)}$, is {\em not a flavor singlet},
and so, as shown above, 
the corresponding density couples to a Goldstone boson,
contrary to naive expectation.
Mathematically, this peculiarity is due to the fact that $\str(I)=0$,
which leads both to the key result that $\str(T_a T_b)$ is not diagonal,
and to the semi-direct product structure of the quenched
symmetry group, \eq{pretendsymmQ}.
This is the sense in which the symmetry structure of the quenched
theory leads to the need to include the $\Phi_0$.

In \ref{app:alter} we give an alternative argument for the need
to keep $\Phi_0$ in quenched, but not in PQ, theories.

\end{subsection}

\end{section}

\begin{section}{Effective Lagrangian}
\label{sec:effL}

We now turn to the construction of
the low energy effective Lagrangian for PQ QCD.
We first recapitulate the standard approach, based on
the ``fake'' symmetry group discussed previously.
We point out the problems with this approach,
and spend most of this section discussing
the extent to which they can be alleviated.
Our conclusion is that the standard approach is appropriate when
doing perturbative calculations, although the justification
for using the effective theory is considerably less rigorous
than for QCD itself.
Some of these points have been discussed previously in 
the random matrix literature, e.g. in Ref.~\cite{verbaarschot},
although in the context of effective theories including $\Phi_0$.
Our aims here are to clarify this work, and to
generalize it to the case at hand in which $\Phi_0$ is absent.
Parts of our discussion are based on an analysis of the
quenched effective Lagrangian given in Ref.~\cite{GSS}.

\subsection{The standard approach}

The fields in the effective theory can be limited to those
describing the pseudo-Goldstone hadrons. These, as we have seen,
are in one-to-one correspondence with the
generators of 
\begin{equation}
[SU(N_V+N|N_V)_L\otimes SU(N_V+N|N_V)_R]/SU(N_V+N|N_V)_V
\,.
\end{equation}
In the standard approach~\cite{BGPQ} 
the pseudo-Goldstone fields lie in this coset space, with the parameterization
\begin{align}
U(x) = \exp\left(\frac{{2i\Phi (x)}}{f}\right),\quad
\Phi (x) = \left( {\begin{matrix}
{\phi (x)}&{\eta _1 (x)}\\
{\eta _2 (x)}&{\tilde \phi (x)}\\
\end{matrix}} \right),
\quad \phi ^\dag   = \phi ,
\quad \tilde \phi ^\dag   = \tilde \phi .
\eqn{usualU}
\end{align}
Here the blocks correspond to quarks and ghosts,
and $\eta_1$ and $\eta_2$ are independent Grassmann matrix fields. 
We are free to interpret $\eta_2$ as $\eta_1^\dagger$, 
in which case $U$ is unitary.
The absence of $\Phi_0$ implies 
the constraint $\str(\Phi)=\tr\phi-\tr\tilde \phi=0$.
The symmetries are implemented as usual:
\begin{equation}
U \to L U R^\dagger\,,\qquad (L,R)\in SU(N_V+N|N_V)
\,,\eqn{LURdag}
\end{equation}
with the VEV $\langle U \rangle=I$ breaking 
the symmetry in the required way 
(note that $U$ transforms like $\widetilde \Omega$---see \eqs{VEV}{VEVtrans}).
The invariant effective Lagrangian is
\begin{equation}
\mathcal{L}\left( U \right) = \frac{{f^2 }}{4}
\str \left( {\partial U\partial U^{\dag}  } \right) - \frac{{f^2 }}{4}
\str \left( {\chi U^{\dag} + U  \chi } \right) + \text{higher orders},
\eqn{usualchL}
\end{equation}
where $\chi$ is proportional to the quark mass matrix, 
$\chi=2 B_0 M$, and $B_0$
and $f$ are unknown parameters.
Clearly this development
mirrors that for QCD step by step.  (Note that 
\eq{usualchL} differs from the form of
the chiral Lagrangian for PQ QCD which is usually quoted: The absence of
$\Phi_0$ means that arbitrary functions of $\Phi_0$ which might multiply
each term in the PQ chiral Lagrangian are absent.)

A problem with this effective Lagrangian becomes apparent when 
developing perturbation theory by expanding $U$ about its VEV.
The ``$\str$'' implies that the fields $\tilde \phi$ have kinetic and
mass terms with the wrong sign, so that $\langle U\rangle=I$
is not a minimum of the action.\footnote{%
Strictly speaking, this comment applies to
fields living entirely within $\tilde\phi$.
However, the anomaly constraint implies that one neutral
meson must have components from both $\phi$ and $\tilde \phi$---
that corresponding to the generator $T_{2N_V +N -1}$ in \ref{app:defs}.
Since $\str(T_{2N_V +N -1}^2)=-1$,
this mixed field also turns out to have a kinetic term with the wrong sign.}
This is dealt with in the standard treatment by simply ignoring the
instability.  A justification for the use of ``wrong sign'' propagators for
some of the mesons is that it implements cancelations which correspond at the
quark level to the desired cancelations between valence quark and ghost
loops.  Actually, once $\Phi_0$ is excluded, the connection between
pseudo-Goldstone propagators and underlying quark and ghost flows is less
clear, so this qualitative justification is less convincing.  Clearly a
better-founded treatment is desirable.

A related concern with the standard approach is that
the chiral symmetry group of the PQ QCD functional integral is not 
$SU(N_V+N|N_V)_L\otimes SU(N_V+N|N_V)_R$,
and the unbroken vector subgroup is not $SU(N_V+N|N_V)$.
As explained in \ref{app:fakesym}, the full group is the subgroup of 
$SL(N_V+N|N_V)_L\otimes SL(N_V+N|N_V)_R$, in  which (using the
nomenclature of Ref.~\cite{deWitt})
the ``bodies'' (non-nilpotent parts) of the left- and right-handed
transformations are related by
\begin{equation}
L = \exp(i\Phi_L) \,,\quad
R = \exp(i\Phi_R) \,,\quad
 (\Phi_L)_{gg}\big|_{\rm body}^\dag =
 (\Phi_R)_{gg}\big|_{\rm body} 
\,,
\eqn{funnyLR}
\end{equation}
(with $\str\(\Phi_{L,R}\)=0$)
while (as explained at the end of \ref{app:fakesym})
the vector group is the subgroup of $SL(N_V+N|N_V)$
in which the body of the ghost-ghost part is unitary:
\begin{equation}
L=R=V = \exp(i\Phi_V) \,,\quad
 (\Phi_V)_{gg}\big|_{\rm body}^\dag =
 (\Phi_V)_{gg}\big|_{\rm body} 
\,.
\eqn{funnyV}
\end{equation}
The coset of these symmetries can be parameterized by
transformations with $L=R^{-1}$, so that
\begin{equation}
L = R^{-1}=A=\exp i \Phi_A \,,\qquad
 (\Phi_A)_{gg}\big|_{\rm body}^\dag =
- (\Phi_A)_{gg}\big|_{\rm body} 
\,.
\eqn{truecoset}
\end{equation}
Note that aside from the condition on the body of
the ghost-ghost block, $\Phi_A$ is an arbitrary, straceless matrix.
One would then expect that the correct
Goldstone fields live in this coset space, i.e.
\beq
\begin{gathered}
{U'(x) = \exp\left(\frac{{2i\Phi '(x)}}{f}\right),\quad 
\Phi '(x) = \left( {\begin{matrix}
{\phi '(x)}&{\eta '_1 (x)}\\
{\eta '_2 (x)}&{i\tilde \phi '(x)}\\
\end{matrix}} \right),}\\
{\left. {\tilde \phi '} \right|_{\text{body}}^\dag   
= \left. {\tilde \phi '} \right|_{\text{body}} ,\quad 
\operatorname{str} (\Phi ') = \operatorname{tr} (\phi ') 
- i\operatorname{tr} (\tilde \phi ') = 0.}\\
\end{gathered}\eqn{naiveGoldstone}
\eeq
(Note the crucial factor of $i$ multiplying $\tilde\phi'$.)
$\phi'$ is an arbitrary matrix of complex c-number fields, constrained only by
the stracelessness condition on $\Phi'$.

It turns out that neither \eq{usualU} nor \eq{naiveGoldstone} is
correct, but before discussing this point it is useful to
understand how the form of the effective Lagrangian would differ
were the Goldstone fields given by \eq{naiveGoldstone}.
This field transforms like $U' \to L U' R^{-1}$, with
$L,R$ given by \eq{funnyLR}. The invariant Lagrangian is
constructed from $U'$, ${U'}^{-1}$, the mass term (a spurion
which transforms like ${U'}^{-1}$), and other sources.
The rules for combining these are in one-to-one correspondence
with those for the usual chiral Lagrangian 
(with $U\leftrightarrow U'$, $U^\dag \leftrightarrow {U'}^{-1}$).
Thus one finds that the most general Lagrangian has
the standard form, (\ref{eq:usualchL}), except that
${U'}^{-1}$ appears rather than $U^{\dagger}$.
We stress that this holds to all orders in the chiral expansion.
The fact that $\sdet U'=1$ implies that there are no additional
$\Phi_0$-like terms.
One might be concerned that the resulting Lagrangian has,
in general, an imaginary part.
This will turn out not to be true in the final form we
consider and so we do not discuss it further here.

\subsection{The correct Goldstone manifold}

Now we return to the issue of the correct Goldstone manifold to use.
Clearly it should be based on the true symmetries,
and thus contained within \eq{naiveGoldstone}.
The problem is that $U'$ in \eq{naiveGoldstone} 
has too many degrees of freedom,
and we must pick an appropriate sub-manifold.
This issue has been addressed by Verbaarschot and collaborators,
in the context of a theory which contains $\Phi_0$ 
(see e.g. Ref.~\cite{verbaarschot}).
They argue, based on the mathematical 
results of Zirnbauer~\cite{Zirnbauer},
that the appropriate integration domain for the effective theory
is the maximal super-Riemannian manifold contained in the coset of the
true symmetry group and the unbroken vector group.
This results in a Goldstone field parameterized by\footnote{%
For explicit non-perturbative calculations, 
the authors of Ref.~\cite{verbaarschot}
use a different parameterization. 
We use the form (\ref{eq:unusualU})
as it is closer to the standard parameterization
of \eq{usualU}.}
\beq
\begin{gathered}
U''(x) = \exp\left(\frac{{2i\Phi ''(x)}}{f}\right),\quad 
\Phi ''(x) = \left( {\begin{matrix}
{\phi ''(x)}&{\eta ''_1 (x)}\\
{\eta ''_2 (x)}&{i\tilde \phi ''(x)}\\
\end{matrix}} \right),\\
{(\phi '')^\dag   = \phi '',\quad (\tilde \phi '')^\dag   = \tilde \phi ''.}\\
\end{gathered}\eqn{unusualU}
\eeq
This form is then inserted in the effective Lagrangian described
in the previous paragraph, i.e.
(\ref{eq:usualchL}) with $U\to U''$
and $U^\dag \to {U''}^{-1}$.
This leads to a Lagrangian whose body is real.

In the following, we discuss the reasoning leading to the form
(\ref{eq:unusualU}), and address a technical difficulty
which arises when extending the argument to the theory without $\Phi_0$.
We then discuss the theoretical foundations of PQ chiral perturbation
theory, and along the way give further physical motivation for
the choice (\ref{eq:unusualU}).

The important differences between the parameterizations (\ref{eq:usualU}),
(\ref{eq:naiveGoldstone}) and (\ref{eq:unusualU}) are in the ghost-ghost
and quark-quark blocks of the exponents.\footnote{%
The anticommuting parts of the exponents
have the same form in all three parameterizations.
We note that the constraint $\eta_1=\eta_2^\dag$ imposed in
(\ref{eq:usualU}) does not change the
rules for Grassmann integration, in which $\eta_1$ and $\eta_2$ are
treated as matrices composed of independent Grassmann variables.
There are also no issues of convergence in the Grassmann integrals.}
We begin by focusing on the former.
Here the difference between the full coset space of \eq{naiveGoldstone}
and the manifold in \eq{unusualU} is
only that the soul of $\tilde\phi'$ is not
required to be hermitian while that of $\tilde\phi''$ is. 
Thus, $\tilde\phi'$ appears to 
contain more degrees of freedom than $\tilde\phi''$.
However, since Gaussian integrals over c-numbers are independent
of the soul, as long as the integrals are convergent 
(see \ref{app:fakesym}),
these extra parameters can be absorbed into $\tilde\phi'$.
Thus the two forms (\ref{eq:naiveGoldstone}) and (\ref{eq:unusualU})
do not differ in this respect.

They do, however, both differ from the standard parameterization, 
\eq{usualU}, by the extra factor of $i$. 
This factor resolves the stability problem raised above. 
If we expand either $U'$ or $U''$ around $I$ to quadratic order,
then the extra factor of $i^2$ cancels the minus sign from the supertrace,
and the kinetic terms for $\tilde\phi'$ or $\tilde\phi''$ (which are equivalent
parameterizations at this order) have the correct sign for 
stability.
This also fixes the overall sign of the kinetic term in the Lagrangian.
The same discussion holds
for the mass term, as long as $M$ is proportional to
the identity. We postpone discussion of the (most relevant) case of
non-degenerate masses until later, since to
address this we need first to understand the
effect of the absence of $\Phi_0$.

We now turn to the quark-quark block of the exponents.
The full coset space (\ref{eq:naiveGoldstone}) contains extra
fields in this block, compared to the standard form
of (\ref{eq:usualU}), since $\phi'$ is not constrained to be
hermitian. These extra fields correspond to the generators
of the non-compact part of the broken symmetry group.
The choice of Ref.~\cite{verbaarschot} is to
keep only the usual, hermitian part of $\phi'$.
A technical reason for doing so is that, if one expands
about $U'=I$, the kinetic term in the action 
(the sign of which is now fixed)
is minimized in these ``hermitian'' directions,
but not in the ``non-compact'' directions.
Thus, with this choice,
the propagators of both $\phi''$ and $\tilde\phi''$
have the correct signs.
For the moment we will accept this as sufficient reason for
making this choice, but return to the point below when
we discuss the foundations of PQ chiral perturbation theory.

\subsection{Constraints from absence of $\Phi_0$}

The analysis so far has not accounted for the constraints
imposed by the absence of $\Phi_0$,
i.e. by the fact that only non-anomalous symmetries should be
represented by the effective Lagrangian.
There is no problem when we consider the full coset space of
\eq{naiveGoldstone}. The constraint is that $\str(\Phi')=0$,
and this can
be satisfied by expanding the diagonal or ``neutral'' part of 
$\Phi'$ using the basis for diagonal straceless generators 
given in \ref{app:defs}:
\begin{equation}
\Phi'_{neu} = \sum_{a=1}^{2N_V+N-1} {\sigma'}^a T_a \,.
\end{equation}
The important point for us is that while all but
one of these generators are contained entirely within either the quark
or the ghost sectors, there is one generator
($T_{2N_V+N-1}$ in our ordering) with components in both sectors.
Now consider the restricted manifold of \eq{unusualU}.
Here the constraint $\str(\Phi'')=\tr\phi''-i\tr\tilde\phi''=0$ 
is made more stringent by the fact that $\phi''$ and 
$\tilde\phi''$ are hermitian. Thus $\phi''$ and $\tilde\phi''$
must be separately traceless, and there can be no component
of $\Phi''_{neu}$ proportional to $T_{2N_V+N-1}$.
Thus the restrictions of the manifold remove not only $\Phi_0$,
but also another neutral generator. This is a problem because,
as seen in the previous section, there are long range correlations
in channels with the quantum numbers of this generator.
In other words, $U''$ is missing one of the neutral 
pseudo-Goldstone bosons.

This problem can be solved by a small change in $U''$.
As we have seen, $U''$ was constructed so that all bosonic fields
have correct sign propagators. For the neutral fields, this 
requires taking components in the quark sector to be real, but
those in the ghost sector to be imaginary. The generator
$T_{2N_V+N-1}$ is problematic because it straddles both sectors.
However, since it satisfies $\str(T_a^2)=-1$, and is thus ghost-like,
we propose that the corresponding field, 
${ \sigma''}^{2N_V+N-1}$, should also be taken imaginary.
Explicitly, we propose replacing $\Phi''$ in eq.~(\ref{eq:unusualU}) with
\beq
\begin{gathered}
\Phi'' = \Phi''_{ch} + \Phi''_{neu}\,, \\
\Phi''_{ch} = \left( {\begin{matrix}
{\phi''_{ch} }&{\eta ''_1 }\\
{\eta ''_2 }&{i\tilde \phi''_{ch}}\\
\end{matrix}} \right),\quad 
{(\phi''_{ch})^\dag   = \phi'',
\quad (\tilde \phi''_{ch})^\dag   = \tilde \phi''_{ch}.}\\
\Phi''_{neu} = \sum_{a=1}^{N_V+N-1} {\sigma''}^a T_a
		+ \sum_{a=N_V+N}^{2N_V+N-1} i {\sigma''}^a T_a \\
\end{gathered}\eqn{finalunusualU}
\eeq
Here ``$ch$'' refers to the off-diagonal, or charged, parts of the field.
In this way we include all the neutral Goldstone flavors,
while maintaining the anomaly constraint, and the condition
that all kinetic terms have the correct sign.

This is not quite the end of the story.  The convergence of the functional
integral depends also on the mass term. Given the choice of
Goldstone fields just described,
the Goldstone boson mass matrix can be calculated, and the 
functional integral
converges only if its real part is positive definite. As
one might expect, there is competition between the sea quark masses and
the valence (and ghost) ones. It turns out (the derivation is somewhat tedious
but straightforward, and we do not include the details here) that the quark 
masses must satisfy
\begin{align}
N_V \overline {\chi _V^{ - 1} } (N_V \overline {\chi _V }  
+ N\overline {\chi _S } ) < (N_V  + N)^2 .\eqn{Mratio}
\end{align}
Here $\overline {\chi _V }$ and $\overline {\chi _S }$ are the average valence
and sea quark masses, and $\overline {\chi _V^{ - 1} }$ the average inverse
valence quark mass.
Restricting lattice simulations to satisfy this constraint is undesirable and
very likely unnecessary. Although the theoretical description of the
simulations in terms of the PQ chiral effective theory is ill defined when
relation \eq{Mratio} is violated, there is nothing special about this point in
parameter space as far as the underlying PQ QCD is concerned. Thus, the
simulations will show no special behavior when \eq{Mratio} is violated. In
addition, chiral perturbation theory is insensitive to these global
convergence considerations, and if {\em it} is regular at this point, 
and provides an adequate description of low energy PQ QCD when the 
quark masses are such that
the inequality is satisfied, it is unlikely that the chiral expressions
derived from this theory will suddenly cease to make sense just when the 
inequality becomes an equality.

Aside from this point, our construction has yielded an effective
Lagrangian with the correct number of Goldstone fields
(i.e. the same number as there are independent Ward identities),
and which one can consistently expand about $\langle U''\rangle=I$.
While in principle one can work with this Lagrangian,
it is more convenient for perturbative calculations to change variables
as follows:
\beq
\begin{gathered}
\tilde\phi''_{ch} = -i \tilde \phi_{ch}\,, \quad
{\sigma''}^{a} = -i \sigma^{a},\ \ a=N_V+N, 2N_V+N-1\,,\\
\eta''_{1,2} = \eta_{1,2} \,,\quad
\phi''_{ch} = \phi_{ch} \,,\quad
{\sigma''}^{a} = \sigma^{a},\ \ a=1, N_V+N-1\,,\\
\end{gathered}
\eeq
The effect of this change is that $U''$ now appears to have
the form of $U$ in eq.~(\ref{eq:usualU}).
Since the underlying integrations are unaltered,
the only effect of this change of variables is to shuffle factors
of $i$ between propagators and vertices (and possibly external fields).
For example, all the 
charged $\tilde\phi$ propagators now have an overall ``wrong'' sign,
while vertex factors appear to satisfy the graded $SU(N_V+N|N_V)$ 
symmetry. Indeed, we can now pretend that $U\in SU(N_V+N|N_V)$,
and ignore the subtleties of this section, as long as we
use the prescription that wrong sign kinetic terms lead to  wrong
sign propagators. 

\subsection{Foundations of PQ chiral perturbation theory}

As promised, we now return to the foundations of PQ chiral perturbation
theory. We first revisit the issue of the extra ``non-compact''
generators in the quark sector. 
Since this sector is a simple generalization
of QCD, the construction and justification of
its effective low energy theory is well understood (see, e.g.
Refs.~\cite{weinberg,leutwyler}). 
In particular, the issue of extra generators
arises also in QCD itself, though it is rarely discussed.

It is useful to recall the steps needed to construct the effective Lagrangian
for QCD-like theories with $N_f$ flavors.  
First, one derives the Ward identities following from
the chiral symmetries.  Second, given the assumed VEV, Goldstone's theorem
determines that there is one light pion field for each axial current.  Third,
one assumes pion dominance of correlation functions, i.e. that the light
fields are the only relevant degrees of freedom.  One then writes down the
most general local Lagrangian incorporating these degrees of freedom which is
invariant under the chiral symmetries, and asserts that this will yield the
most general amplitudes consistent with the Ward identities and the principles
of relativistic quantum mechanics~\cite{weinberg}.  To formalize this, one
introduces sources for currents, scalar and pseudoscalar densities, etc., and
defines the generating functional in the usual way
\begin{align}
Z_{\text{QCD}}
[f_L^\mu ,f_R^\mu ,S,P, \ldots ] = 
1 + \int
{\left\langle {j_\mu ^L (x)} \right\rangle f_L^\mu (x)} + \ldots 
\end{align}
The Ward identities of the theory are encapsulated in the
invariance of $Z$ under gauge-like transformations,
\begin{equation}
\begin{split}
f_L^\mu &\to L f_L^\mu L^{-1} -i [\partial^\mu L)] L^{-1} \,,\ \
f_R^\mu \to R f_R^\mu R^{-1} -i [\partial^\mu R)] R^{-1} \,,\\
S-iP &\to L(x) [S - iP ] R^{-1}(x)\,,\ \
S+iP \to R(x) [S + iP ] L^{-1}(x)\,.\\
\end{split}
\eqn{gauge}
\end{equation}
A similar definition is used for
$Z_{\text{eff}} [f_L^\mu ,f_R^\mu ,S,P, \ldots ]$.
The claim is that,
by adjusting the coefficients in the effective Lagrangian, including
contact terms, one can match the generating functional between QCD and the
effective theory,
\begin{equation}
\eqn{ZeqZ}
Z_{\rm eff}[f_L^\mu,f_R^\mu,S,P,\dots]
= 
Z_{\rm QCD}[f_L^\mu,f_R^\mu,S,P,\dots]
\,,
\end{equation}
to any desired accuracy in a momentum and quark mass 
expansion~\cite{leutwyler}.

The issue at hand is what group the transformations $L$ and $R$
belong to. This depends on whether $Z_{\rm QCD}$ is defined using
the operator formulation or with a functional integral.
In the former case,
the transformations of $\psi$ and $\psi^\dagger$ are related,
since $\bar\psi= \psi^\dagger \gamma_0$.
Thus only unitary chiral transformations are symmetries
[as in \eq{Utrans}], and
$L(x)$ and $R(x)$ in (\ref{eq:gauge}) are elements of $SU(N_f)$. 
On the other hand, in the functional integral formulation,
the fact that $\psi$ and $\bar\psi$ are independent variables leads
to a larger symmetry (see \ref{app:fakesym}): $L,R\in SL(N_f)$.  
Which of these symmetries should $Z_{\rm eff}$ respect?  
The answer is the smaller, unitary symmetry.
This can be seen in two ways. First, the ``extra''
transformations contained in $SL(N_f)/SU(N_f)$ lead to non-hermitian sources
$S$ and $P$ [see \eq{gauge}], 
and so to a non-hermitian Hamiltonian. Thus they move us out of
the space of physical theories.  Second, as shown in
 \ref{app:fakesym}, the extra transformations do not lead to additional Ward
identities, and thus do not lead to additional Goldstone bosons.  Since the
effective Lagrangian should certainly respect the symmetries of the operator
formulation of the theory, and this leads to all the desired Goldstone bosons,
one should not extend the effective theory to include non-compact
symmetries.
%
%
We conclude that the ``normal'' choice of pion
fields in the quark block of the Goldstone matrix, eq.~(\ref{eq:unusualU}), 
is correct. The only exception is for the part of $\phi''$
proportional to the identity
in this block (i.e. the field ${\sigma''}^{2N_V+N-1}$).
The argument just given does not apply to this field since it is
a flavor singlet in the quark block.

Finally, we address the extent  to which one  can derive the effective
Lagrangian  in the  graded  sector of   the  theory  (i.e.  the  parts
involving ghosts). The first two steps followed in the quark sector go
through in this sector as well: the Ward identities of  PQ QCD are the
graded  generalizations  of those in  QCD,   and the symmetry breaking
pattern in PQ QCD  follows from that  in  QCD. From these  results, we
have established the presence of massless poles in two-point functions
of   broken symmetry generators.  In  the  standard approach, the next
step  is to interpret   each pole as being   due to a  physical single
particle   state created  by  the corresponding  operator,  from which
follows the result   that any  correlation function including    these
operators will have poles at the same positions.  This deduction is
key in justifying the effective  chiral Lagrangian for QCD, but  its
extension to PQ QCD is  not obvious.   Furthermore, when one  includes
quark masses as small perturbations in QCD, one concludes from similar
arguments that  this results in  a small shift  in the position of the
poles. In PQ QCD, on the other hand, quark masses have a substantially
different effect--they can lead to the appearance of a double pole at
$p^2=0$ with a coefficient proportional to the quark masses,
as will be seen explicitly in sec.~\ref{sec:poles}.

What is lacking in the PQ case  is a positive-norm Hilbert space 
interpretation, since the theory contains ghosts. 
This is not to say that a derivation of the effective Lagrangian
is not possible for PQ theories, but rather that a generalization
of the standard methods is needed. 
Since 
quark and ghost correlation functions differ simply by signs,
we speculate that the appropriate theoretical framework only differs
from the standard one by requiring a
vector space with indefinite metric (e.g. single anti-ghost states
having negative norm), as well as bosonic ghost creation and 
annihilation operators.
The Hamiltonian would still be physical---having real
eigenvalues bounded from below---and ghost operator commutators
would be causal.
In this way Lorentz invariance is maintained, while unitarity is
lost because of transitions between positive 
and negative norm states.\footnote{%
This  is indeed the  structure  observed  in explicit calculations  of
scattering  amplitudes  in quenched  chiral perturbation  theory:  the
results are Lorentz invariant but not unitary~\cite{colangelo}.} 
It  is perhaps  plausible in such  a  set-up that a generalization  of
standard ``pologogy''   could  be derived,   in   turn leading  to  an
effective Lagrangian, and that  the requirement of Lorentz  invariance
would force this Lagrangian to be local.

\end{section}

\begin{section}{Integrating out the $\Phi_0$}
\label{sec:phi0}

The work of the previous section has shown that we can write the effective
chiral Lagrangian for the PQ theory in terms of an $SU(N_V+N|N_V)$ field $U$,
transforming as in \eq{LURdag}. In this section we need only write the form of
this Lagrangian schematically: 
\beq 
\mathcal{L}(U) = \sum\limits_i {\ell_i O_i(U)} \,, \eqn{Lform} 
\eeq 
where the $\ell_i$ are unknown parameters, and $O_i$
are operators constructed from $U,\ U^\dag$, their derivatives, the rescaled
mass matrix $\chi$, and sources for external operators.  We note that the
operators allowed by the graded chiral symmetry are identical in form to those
allowed by the usual chiral symmetry in QCD. In particular, there are no
additional operators.

An important property of PQ chiral perturbation theory is that correlation
functions involving external sources restricted to lie within the sea-quark
sector are identical to those obtained using the effective chiral Lagrangian
for unquenched QCD (with $N_f=N$, and without the $\eta'$).  This identity,
which is trivial at the quark level, can be seen diagram by diagram in chiral
perturbation theory, due to cancelations between diagrams in which a valence
quark is replaced with a ghost quark.  A compelling general argument in
support of this fact, though not a proof, is the following. Let $\mcal{M}$ be
the mass matrix for a specific valence quark and the corresponding ghost,
$\mathcal{M} = \operatorname{diag} (m_V ,m_V )$.  Let $A$ be an operator which
does not depend on this valence-ghost pair. We write its expectation value as
$\left\langle A \right\rangle = a(\mathcal{M})$, making only the $\mcal{M}$
dependence explicit. The flavor symmetry group includes
$\operatorname{SU} (1|1)$
transformations that mix the valence and ghost quark fields. These
transformations are equivalent to a change of variables in the functional
integral, in addition to a group transformation (by conjugation) of
$\mcal{M}$. Since $A$ is unchanged by the change of variables, it follows that
$a(\mcal{M})$ can depend only on 
$\operatorname{SU} (1|1)$
invariants constructed from $\mcal{M}$. 
These, however, can only be supertraces of powers of the mass matrix
$\operatorname{str} (\mathcal{M}^n) = 0,$
so $a$ is independent of the valence (and ghost) quark mass. To show
independence of $a$ on the {\em existence} of the valence-ghost fields in the
theory one still needs to show that any effect from these fields introduces
$\mcal{M}$ dependence.

It follows that, with external sea-sector sources, one can restrict
the internal $U$ field to take the following block-diagonal form
\beq
U_{\rm QCD} = \left(I_{N_V}, SU(N), I_{N_V} \right)
\,.
\eqn{UQinPQ}
\eeq
Since the terms in $\mcal{L}(U)$ are in one-to-one correspondence
with those in the QCD effective Lagrangian, 
and the restriction (\ref{eq:UQinPQ}) does not change the form of the 
operators,\footnote{%
With \eq{UQinPQ}, $\chi$, $U$ and $\partial U$ are proportional 
to matrices of the general form
\begin{align}
\left( {\begin{matrix}
A&0&0\\
0&B&0\\
0&0&A\\
\end{matrix}} \right),
\end{align}
where the three entries correspond to valence, sea, and ghosts, and $A,\ B$
stand for any $N_V\times N_V$ and $N \times N$ blocks, respectively. Any
product of such matrices still has this structure, which causes each
supertrace in the Lagrangian to reduce to a simple trace over the sea-sea
block.} 
it follows that both PQ and unquenched Lagrangians share the
parameters $\ell_i$.  As we emphasized in \cite{shshI,shshII}, this means that
the parameters $\ell_i$, which describe the chiral expansion for the
unphysical PQ theory, are in fact physical.  For example, at NLO, $\ell_i$ are
the GL coefficients which encode our current experimental knowledge (and
ignorance) of QCD at low energies for the light mesons.
We stress that the $\ell_i$ do depend on $N$, i.e. the GL parameters that
one obtains depend on the number of light sea quarks.

{} From a theoretical point of view, this completes the construction of the PQ
chiral effective Lagrangian, and the demonstration that it contains only
physical parameters. When doing perturbative calculations with this Lagrangian
one must, however, implement the constraint that $\sdet\; U = 1$, or, if we
write $U=\exp(2i \Phi/f)$, that $\str(\Phi)=0$. A standard way of achieving
this is using straceless generators.  For the realistic case of $N_f=3$ and
considering only mesons, $N_V=2$, calculations involve the generators of
$SU(5|2)$, and can be quite tedious.  In this section we point out that a
simple alternative is to reintroduce the $\Phi_0$ field---as a calculational
device and not as a physical field---and then to integrate it out. This allows
one to work in the ``quark basis'', rather than with the actual
pseudo-Goldstone fields.  Calculations in the quark basis are more transparent
since one can trace quark flow through each diagram, and see the cancelations
between valence quarks and ghosts very simply \cite{BGQ,SSchirallogs}.
Keeping $\Phi_0$ also allows us to reinterpret previous calculations, which
have included it as a physical field, as applying to the theory without this
field (a point discussed at the end of this section).

\subsection{Functional integral approach}
We reintroduce $\Phi_0$ by enlarging $U$ to 
$\Sigma \in U(N_V+N|N_V)$,
\beq
\Sigma(x) = U(x) \exp\left(\frac{2 i \Phi_0(x)}{f\sqrt{N}} \right)
\,,
\eeq
and considering a theory with Lagrangian
\beq
\mathcal{L}'(\Sigma )& = & m_0^2 \Phi _0^2  + \mathcal{L}(\Sigma )
\\
{} & = & m_0^2 \Phi _0^2  + \sum\limits_i {\ell_i O_i (\Sigma )}
\,. \eqn{L'aux}
\eeq
In words, we simply replace $U$ with $\Sigma$ in all the terms in
$\mcal{L}(U)$ and add a mass term for $\Phi_0$.
It is useful to make the $\Phi_0$ dependence explicit by expanding
$\Sigma$:
\beq
\mathcal{L}(\Sigma ) =
\mathcal{L}(U) + \sum\limits_{j = 1}^\infty  {R_j(U,\partial)(\Phi_0)^j } 
\,,
\eeq
where we allow $R_j(U,\partial)$ to contain derivatives 
acting on the $\Phi_0$ field. The enlarged theory is then
\beq
{Z}&=& \int 
{D\Sigma \exp \left( { - \int {\mathcal{L}'(\Sigma )} } \right)} \\
&=& \int {DUD\Phi _0 \exp \left\{ { - \int {\left( {\mathcal{L}(U) + m_0^2
\Phi _0^2  + \sum_j {R_j(U) (\Phi_0)^j } } \right)} } \right\}}
\,.
\eeq
We assume that the theory is regulated with a chirally invariant fixed
cut-off, such as the lattice~\cite{CHPTonlattice}.

We now take the limit $m_0\to\infty$, and argue that we then 
return to the original theory with $\Sigma\to U$, i.e. the theory
that we want to do calculations with.
The argument goes as follows. 
We expect the fluctuations in $\Phi_0$ to be
of $\mathcal{O}(1/m_0 )$, and thus that all the $R_i$ terms,
which are independent of $m_0$, are suppressed compared to the mass term. 
Similarly, correlation functions calculated in this 
theory should become independent of $R_i$ at $m_0\to\infty$.
Furthermore, the expectation value of any operator
$O(\Sigma ) = O(U) + \sum_j {r_j (U)(\Phi_0)^j } $ satisfies
\beq
\langle O(\Sigma )\rangle_{\mathcal{L}'} 
&= & \frac{1}{Z_{\mathcal{L}'}(m_0)}\int\! {DUD\Phi_0 \,O(\Sigma )
\exp \left( { \!-\! \int\! {\mathcal{L}'(\Sigma )} } \right)} \\
&\xrightarrow[{m_0  \to \infty }]{}& 
\frac{1}{Z_{\mathcal{L}'}(\infty)}\int\! {DUD\Phi _0 \,O(U)
\exp \left\{ { \!-\!\int\! {(\mathcal{L}(U) + m_0^2 \Phi _0^2 )} } \right\}} \\
& =& \frac{1}{Z_{\mathcal{L}}}
\int\! {DU\,O(U)\exp \left( { \!-\! \int\! {\mathcal{L}(U)} } \right)}\\
&=& \langle O(U)\rangle_\mathcal{L} 
\,.
\eeq
So we obtain correlation functions in the theory we want,
with any $\Phi_0$ contribution to the external
operators being projected out.

It is crucial for this argument that the theory be regulated in such
a way that loop momenta are limited.\footnote{We thank David B. Kaplan
for emphasizing this point to us.}
This allows $\Phi_0$ to be integrated out in a trivial way.
Without a fixed cut-off, the $(\partial \Phi_0)^2$ term
[implicitly contained in $R_2(U) \Phi_0^2$] 
can dominate over the $m_0^2$ term in loop integrals. 
This in turn leads to a non-decoupling of $\Phi_0$.
For example, using dimensional regularization,
$\Phi_0$ tadpole diagrams give contributions proportional to
$(m_0^2/f^2) \ln(m_0/\mu_{\rm DR})$. With a fixed cut-off, by contrast,
the contribution is of the form $\Lambda^4/(f^2 m_0^2)$, and vanishes
when $m_0\to \infty$.

There is a second subtlety which could invalidate the argument just
presented. In physical theories heavy particles ``decouple'', meaning that
physics at energies much lower than their mass is independent of the details
of their interactions and dynamics. This idea is at the core of effective
field theories in which only light particles are included as explicit
dynamical degrees of freedom. In this sense, taking the mass of a particle to
infinity in physical theories is well defined.  For unphysical theories,
however, one might be concerned that this limit does not exist.  That this is
a legitimate concern is shown by the fact, explained below, that the limit
cannot be taken in quenched QCD.  We will see, however, that this
is not a problem for the PQ theory.

\subsection{$\Phi_0$ in the flavor-neutral propagator}
To understand this point, and also to gain insight into the nature of the
$m_0\to\infty$ limit, we consider the form of the propagator of ``neutral''
mesons, i.e. those created by the diagonal elements of $\Phi$. It is
sufficient to consider this propagator since this is the only place that the
$m_0^2$ term enters when one develops perturbation theory.  We begin with some
notation. We use $\Pi(x)$ to denote the pion field including $\Phi_0$: 
\beq
\Sigma = \exp\left(\frac{2i \Pi}{f}\right) \,,\qquad \Pi(x) = \Phi(x) +
\Phi_0(x) \frac{I}{\sqrt{N}} \,.  
\eeq 
The set of meson fields $\{\pi_{ij}\}$
that make up the quark basis is defined through
\begin{align}
\Pi (x) = \sum\limits_{i,j = 1}^{2N_V  + N} {T_{ij} \pi ^{ij} (x)} ,
\quad (T_{ij} )_{kl}  = \delta _{ik} \delta _{jl} ,
\end{align}
and the flavor neutral part is therefore
\begin{align}
\Pi _{\text{neu}} (x) = 
\sum\limits_{i = 1}^{2N_V  + N} {T_{ii} \pi ^{ii} (x)} .
\end{align}
The neutral part of $\Pi$ can also be decomposed using the 
basis of generators of $U(N_V+N|N_V)$ given in \ref{app:defs}:
\beq
\Pi_{\rm neu}(x)
= \sum_{a=1}^{2N_V+N} T_a \sigma^a(x)
\,,
\eeq
where the generators satisfy
\beq
\str\(T_a T_b\) = g_{ab}.
\eqn{gab}
\eeq
The first $2N_V+N-1$ components of $\sigma^a$ are true
pseudo-Goldstone fields, while the last entry is the field
added by hand, $\Phi_0$. 

We consider all the quadratic terms in the action that contribute to the
leading order propagator. At this point we do not specify what these terms
are, only require that the $\Phi_0$ mass term is included.  For a given
momentum, the inverse propagator will have the following form:
\begin{align}
G_{(\sigma )}^{ - 1}  = \left( {\begin{matrix}
A&B\\
C&{m_0^2  + d}\\
\end{matrix}} \right).\eqn{GsigInv}
\end{align}
Here we have separated off the last row and column of the matrix,
corresponding to entries involving $\Phi_0$. Thus $A$ 
[a $(2N_V+N-1)\times(2N_V+N-1)$ matrix] is the
inverse propagator in the theory without $\Phi_0$.
The important point is that $A$, $B$, $C$ and $d$ are independent of
$m_0$, and finite (because of the momentum cut-off).
Using this, one can easily show (\ref{app:Ginverse}) that the only
requirement needed for the propagator to have a limit as $m_0\to\infty$ 
is that $A$ be non-singular.
The limit is then
\begin{align}
\overline G_{(\sigma )}  = 
\mathop {\lim }\limits_{m_0  \to \infty } G_{(\sigma )}  
= \left( {\begin{matrix}
{A^{ - 1} }&0\\
0&0\\
\end{matrix}} \right).
\end{align}
This propagator has the two properties needed so that one is effectively doing
the calculation using only the physical $SU(N_V+N|N_V)$ degrees of freedom:
(1) its projective form removes $\Phi_0$ from the theory---factors of $\Phi_0$
in vertices and external fields simply do not propagate. 
(2) the
propagator in the physical subspace, $A^{-1}$, is the correct propagator for
these degrees of freedom alone.
\footnote{
One might be concerned that the fact that some of the fields
have the wrong metric might lead to a basis dependence
in the results of calculations. Indeed, the transformation
between the ``sigma'' and ``pi'' bases,
$\sigma^a =\sum_b{\Lambda^{a}}_b \pi^{bb}$,
is not unitary, but instead is a generalized Lorentz rotation.
This follows from the results
\beq
\str \( \Pi_{neu}^2\) = \sum_{a,b}\sigma^a g_{ab} \sigma^b
= \sum_{a,b}\pi^{aa} \tilde g_{ab} \pi^{bb} \,,
\eeq
where $g_{ab}$ (defined in \ref{app:defs}) plays the role of the metric 
tensor, and $\tilde g_{ab}$ has the same signature but with elements permuted.
Since Lorentz-like rotations do not preserve Euclidean inner products,
projections onto subspaces are not in general
invariant under these rotations.
This turns out, however, not to be relevant in our calculations.
All that we need in order to show the basis independence of our calculations
is that the propagator of the matrix field,
$\langle \Pi_{neu}(x) \Pi_{neu}(0)\rangle$,
which is the building block of perturbation theory,
is the same in either basis. It is straightforward to show
that for this to hold, it is sufficient for $\Lambda$ to be invertible,
which is clearly the case.}

Thus the only remaining question is whether $A^{-1}$ exists.  To study this we
first consider only the kinetic + $m_0^2$ terms in the Lagrangian
\eq{L'aux}. $A$ can be easily read off, using 
\beq 
\str\(\partial \Pi_{neu} \partial \Pi_{neu} \) \propto 
g_{ab} \partial \sigma^a \partial \sigma^b \,,
\eqn{eq:PQkin} 
\eeq 
and the explicit form of $g_{ab}$ from \ref{app:defs}:
\begin{align}
\begin{split}
{\text{PQ}}&:
{\quad A = p^2 \operatorname{diag} (1, \ldots ,1, - 1, \ldots , - 1, - 1)}\\
\text{Q}&:
{\quad A = p^2 \operatorname{diag} (1, \ldots ,1, - 1, \ldots , - 1,0)}\\
\end{split}\eqn{PQD}
\end{align}
Clearly $A$ is invertible only in the PQ case, and the limit $m_0\to\infty$
can be taken only in that theory, but not in quenched QCD.

The usual mass term in the Lagrangian, which is of the same order in chiral
perturbation theory as the kinetic term, can be treated as a vertex. This
leads to a geometric series of tree diagrams, all contributing to the leading
order. The propagator in these diagrams, when it exists in the $m_0\to\infty$
limit, removes any $\Phi_0$ contributions. This means that the conclusions
derived above from eqs.~(\ref{eq:PQD}) still hold. There is, however, a
potential loophole in this argument: the infinite sum over tree level diagrams
and the $m_0\to\infty$ limit might not commute. That this is not a problem can
be seen directly from previous calculations of the LO propagator in chiral
perturbation theory in which the quark mass term was included before taking
the $m_0\to\infty$ limit.  In the case of PQ QCD with 3 sea quarks, we
have~\cite{BGPQ,shshII}
\begin{align}
\begin{split}
{G_{AA}^{neu} }& \equiv {\int {d^4 xe^{ - ip \cdot x} 
\left\langle {\pi _{AA} (x)\pi _{AA} (0)} \right\rangle } }\\
{}& = {\frac{1}
{{p^2  + \chi _A }} - \frac{{(p^2  + \chi _1 )
(p^2  + \chi _2 )(p^2  + \chi _3 )(m_0^2 /3)}}
{{(p^2  + \chi _A )^2 (p^2  + M_{\pi _0 }^2 )
(p^2  + M_\eta ^2 )(p^2  + M_{\eta '}^2 )}}}\\
{}&{\xrightarrow[{m_0  \to \infty }]{}}{\frac{1}
{{p^2  + \chi _A }} - \frac{1}
{3}\frac{{(p^2  + \chi _1 )(p^2  + \chi _2 )(p^2  + \chi _3 )}}
{{(p^2  + \chi _A )^2 (p^2  + M_{\pi _0 }^2 )(p^2  + M_\eta ^2 )}}}\\
\end{split}
\end{align}
where $A$ is a valence quark label, $\chi_{i}$ normalized quark masses (with
$i=1,2,3$ for sea quarks), and $M_{\pi _0 },\ M_\eta,\ M_{\eta'} $ masses of
neutral mesons in the sea sector (%
$M_{\eta '}^2  = m_0^2  + \mathcal{O}(\chi_i /m_0^2 )$
).
Clearly the limit $m_0\to\infty$ exists, so the potential problem
does not arise.

The corresponding result in the quenched theory is
\begin{align}
G_{AA}^{neu}  = \frac{1}
{{p^2  + \chi _A }} - \frac{{m_0^2 /3}}
{{(p^2  + \chi _A )^2 }},
\end{align}
which shows that the $m_0\to\infty$ limit cannot be taken,
irrespective of the quark mass.
The fact that one {\em cannot} integrate out $\Phi_0$ in the quenched theory
is consistent with the result, discussed in section \ref{sec:lowpoles},
that $\Phi_0$ {\em should not} 
be projected out of the theory, as it contributes to
long range correlations. 
We observe again the central role played by the structure of 
$\operatorname{str} (T_a T_b )$
in the analysis.

\subsection{Reinterpreting previous calculations}

We have argued above that the $\Phi_0$ field need not be included
in PQ chiral perturbation theory because, in the massless limit,
its propagator does not have a massless Goldstone pole,
but instead has a pole at the $\eta'$ mass (if $N=3$).
Since chiral perturbation theory does not converge for 
$|p|\approx 1{\rm GeV}$,
it will not, in general, be useful for $|p|\approx m_{\eta'}$.
Thus the natural choice is leave $\Phi_0$ out of the effective
Lagrangian describing partially quenched simulations of QCD,
perhaps adding it back as a mathematical device to simplify calculations,
as described previously in this section.

Previous calculations in PQ chiral perturbation theory,
e.g. Refs.~\cite{BGPQ,shZ,GLPQ,shshII,GP}, have,
however, included $\Phi_0$ as a physical field, 
rather than as a mathematical device. 
The reasons for this are partly historical 
(PQ chiral perturbation theory is an extension of quenched chiral perturbation
theory, in which $\Phi_0$ must be kept),
and partly motivated by physics (there are regimes, e.g. large $N_c$
or possibly $N=2$, in which $\Phi_0$ is lighter than $\Lambda_\chi$, 
and should
be treated differently from other non-Goldstone hadrons---see also the
discussion in Ref.~\cite{GLPQ}). 
Our purpose here is
 to show that these previous calculations
can be used to obtain the results in the theory without $\Phi_0$.

We recall that including $\Phi_0$ as a
physical field introduces several problems.
First, chiral power counting is lost---adding a loop
does not lead to an extra small factor of size $m_\pi^2/\Lambda_\chi^2$,
but rather to a factor of size $m_0^2/\Lambda_\chi^2$.
If this latter factor is not small, diagrams with any number of $\Phi_0$
loops must be included.
Second, because of the anomaly, one can multiply terms in
the effective Lagrangian by arbitrary functions of $\Phi_0$
consistent with parity invariance. 
This introduces many new and poorly known parameters.
Finally, the parameters $\ell_i$ are those of
unquenched chiral pertubation theory including the $\eta'$,
which are related non-perturbatively, in a poorly known way, to those
of the usual QCD chiral Lagrangian without the $\eta'$.
It is possible that one can mitigate some of these problems by
assuming that the additional coupling constants are small
(since they are $1/N_c$ suppressed).\footnote{Indeed, there has been
significant success in applying quenched chiral perturbation
theory to lattice data, despite the presence of the same 
problems (see, for example, Ref.~\cite{thacker}).}
Such an analysis, however, inevitably involves assumptions
which go beyond the systematic application of chiral perturbation theory.

What we note here is that if one takes the limit $m_0\to\infty$
in the results of these calculations, one recovers
the results in the theory without $\Phi_0$ in which the
parameters are those of the QCD chiral Lagrangian without the $\eta'$.
The argument is a simple extension of that given in the previous section.
If we expand in powers of $\Phi_0$, the total Lagrangian is 
\beq
\mathcal{L''}(\Sigma ) = m_0^2 \Phi_0^2 +
\mathcal{L}(U) + \sum\limits_{j = 1}^\infty  {R'_j(U,\partial)(\Phi_0)^j } 
\,,
\eeq
This differs from \eq{L'aux} only in the replacement of the
$R_j(U,\partial)$ terms by new functions $R'_j(U,\partial)$ which include
the additional $\Phi_0$ couplings. For example,
a term proportional to $\alpha (\partial \Phi_0)^2$ is usually
included. However, since the
$R'_j$ do not depend on $m_0$, the discussion of the previous
section shows that they are irrelevant as $m_0\to\infty$.
The remainder of the argument continues as before.

What happens to the previous calculations when $m_0\to\infty$?
First, the neutral propagator goes to an $m_0$ independent
limit---the same limit discussed in the previous section---including, 
in general, light single and double poles.
Second, ``$\eta'$ loops''---those involving propagators 
with poles at $m_\eta'\sim m_0$---give vanishing contributions.
Third, all additional $\Phi_0$ couplings, such as $\alpha$, 
do not contribute---as
argued on general principles above.
Thus the results of the calculations simplify considerably.

In fact, this simplification has been noted and used previously,
though not fully justified \cite{shZ,shPQ,GLPQ,GP}. It has been argued that,
because $m_0\sim \Lambda_\chi$, terms proportional to
$m_\pi^2/m_0^2$ are higher order in chiral perturbation
theory and contribute only at the two-loop level.
They can thus be dropped in NLO calculations.
Contributions from $\eta'$ loops are not higher order
(as noted above they can grow as $\ln m_0$ in dimensional
regularization), but have been dropped by hand because they would not be
present in QCD chiral perturbation theory calculations without the $\eta'$.
Together, this amounts to integrating out $\Phi_0$ perturbatively, 
whereas a non-perturbative
treatment is required. Our results effectively provide such a treatment,
and justify the ad-hoc procedures adopted previously.

\end{section}

\begin{section}{Structure of the flavor-singlet propagator}
\label{sec:poles}

A signature prediction of PQ chiral perturbation theory at LO and NLO concerns
the peculiar features of the pole structure of correlation functions
(``polology'' for short).  The effective theory predicts that propagators of
flavor diagonal (``neutral'') mesons have both single and double pole
singularities. In this section we show that, with some plausible assumptions
about ``normal'' aspects of PQ polology, the occurrence of the unphysical
double poles is a consequence of the symmetry structure of the theory. It
follows that the double poles really are a feature of PQ QCD, and the grounds
for this important prediction of PQ chiral perturbation theory are better
established and understood.  We first present the argument in terms of quark
fields, so that it applies directly to PQ QCD, and then point out how the
argument extends to the effective theory.

\begin{subsection}{PQ QCD}

For simplicity, we discuss only the case of
degenerate sea quarks.

Consider the two-point correlator of pseudoscalar quark bilinears,
\begin{equation}
G_{ijkl}(x) = \left\langle {\Pi_{ij}(x) 
\Pi_{kl}(0)} \right\rangle 
\,,\qquad
\Pi_{ij} = \tr\left( Q_i \overline Q_j \gamma_5\right) \,.\eqn{Cdef}
\end{equation}
The trace is over Dirac and color indices, and the order of $Q$ and 
$\overline Q$ in the bilinears
is chosen to simplify the transformation properties.
If $i=j$ and $k=l$, we call the correlators ``neutral'', while
if  $i\ne j$ and $k\ne l$ we call them charged.\footnote{%
In the latter case,
the unbroken vector symmetries require that $i=l$ and $j=k$.
This can also be seen by evaluating the correlator in terms of
quark propagators.}

We first focus on the neutral propagator in a particularly simple
theory, that with a single valence quark, $A$, its
ghost, $\tilde A$, and one sea quark, $S$.
In ~\ref{app:Gstructure} we derive the constraints on $G$ that
follow from the graded vector symmetry together with
Euclidian translation and rotation invariance.
We find that the neutral propagator $G_{jjkk}$
takes the following form in the
$(A\overline{A},S\overline{S},\tilde{A}\overline{\tilde{A}})$ basis:
\begin{equation}
G = \left( {\begin{matrix}
{r + s}&t&r\\
t&u&t\\
r&t&{r - s}\\
\end{matrix}} \right)\eqn{Gstruc}.
\end{equation}
The form holds at all separations and thus also in momentum space.
By following quark propagators,
we can interpret the elements of $G$ as follows.
(We use the language of perturbation theory, although the results
hold non-perturbatively.)
The off-diagonal term $G_{AA\tilde{A}\tilde{A}}$ receives
contributions only from disconnected diagrams,
because the quark lines carrying
the $A$ flavor cannot be contracted with the $\tilde{A}$ flavors.  
$G_{AAAA}$, on the other hand, 
can get contributions from both connected {\em and}
disconnected diagrams. All the disconnected graphs contributing to either of
these terms in $G$
have exactly the same structure, except for the interchange of
$\tilde{A}$ propagators with $A$ ones. 
Since these propagators are equal, and there
are no relative signs from Wick contractions,
it follows that $r$ in \eq{Gstruc} is the sum of all disconnected
diagrams, and $s$ is the sum of all connected ones.\footnote{%
One can also show, using the graded symmetries,
that $s$ is related to correlators which are clearly charged:
$s = G_{\tilde A A A \tilde A } $
and $s = G_{ABBA}$ if $m_A=m_B$ (see \ref{app:Gstructure}).}
Similarly, $t$ is the sum of all disconnected diagrams but with
the quark line coupled to one of the external operators
replaced by a sea quark (so that $t=r$ if $m_A=m_S$).
Finally, $u$ is the complete $S\overline{S}$ propagator,
including both connected and disconnected contributions.
Note that the relative signs of the various contributions can be
determined simply by considering Wick contractions.

Turning now to the more interesting case of $N>1$ degenerate 
sea quarks, it is straightforward to show that the same
form of $G$ holds for the neutral propagator in the
basis in which $S\overline{S}$ is replaced by 
$\eta' \propto \sum_{j=1,N} S_j \overline{S_j}$.
The quantities $u$, $r$, $s$, and $t$ are of course different,
with $u$ being the $\eta'$ propagator.
In this theory, there are $N-1$ additional neutral bilinears,
but their correlators with $A\overline{A}$, $\tilde A\overline{\tilde A}$
and $\eta'$ vanish since the latter are invariants under any symmetry 
transformation that involves only sea quarks, while none of the former is.
Thus $G$ is block-diagonal and we can consistently consider only
the $3\times 3$ block \eq{Gstruc}.

For future use, we separate out from $G$ the quantity
\begin{equation}
G_C  \equiv \operatorname{diag} (s,u, - s) \,.
\end{equation}
We do this because we expect these quantities to have standard pole structure.
The $\eta'$ propagator $u$ is physical, and so has only single poles, 
with none of them being light. 
As for the connected propagator $s$, 
our key assumption in this section is that
it has a light single-pole. This is what is predicted by PQ chiral
perturbation theory at LO and NLO \cite{shshII}, 
and also what is observed in numerical
simulations of PQ theories.

In order to better understand the properties of the
elements of $G$, it is useful to study the form of its inverse:
\begin{equation}
G^{ - 1}  = \left( {\begin{matrix}
{R + S}&T&{ - R}\\
T&U&{ - T}\\
{ - R}&{ - T}&{R - S}\\
\end{matrix}} \right)\eqn{GIstruc}
\end{equation}
with
\begin{eqnarray}
&& R =  - \frac{{ur - t^2 }}{{us^2 }}
;\quad S = \frac{1}{s}
;\quad T =  - \frac{t}{{us}}
;\quad U = \frac{1}{u} 
\\ 
&& r =  - \frac{{UR - T^2 }}{{US^2 }}
;\quad s = \frac{1}{S}
;\quad t =  - \frac{T}{{US}}
;\quad u = \frac{1}{U}\,.
\end{eqnarray}
We now decompose $G^{-1}$ as
\begin{equation}
G^{ - 1}  = G_C^{ - 1}  + \Sigma\,, \eqn{GI}
\end{equation}
where 
\begin{equation}
  \Sigma   = \left( {\begin{matrix}
R&T&{ - R}\\
T&0&{ - T}\\
{ - R}&{ - T}&R\\
\end{matrix}} \right)\,.
\end{equation}
The form of \eq{GI} is the same as that in perturbation theory,
with $G_C$ corresponding to the free propagator, 
and $\Sigma$ to the one-particle irreducible self-energy.
Thus we interpret $\Sigma$ as the self-energy
contributions arising from disconnected diagrams involving
at least some valence quarks or ghosts
(disconnected contributions involving sea quarks having
already been included).

We can now state our second assumption: that the elements
of $\Sigma$ ($R$ and $T$)
are regular in momentum space at the position of the light pole in $s$.
Our reasoning here is that $\Sigma$ is one-particle irreducible
with respect to the connected propagator, and thus does not have
simple light poles. The only source for non-analyticity at the position
of the light pole would be a deeply bound state of pseudo-Goldstone bosons,
but this should not be present because the interactions are weak.

Given our assumptions, we can now read off the pole stucture of $G$,
by expressing it in terms of quantities with known,
or assumed, behavior (simple light poles in $s$, around which $u,\ R$, 
and $T$ are regular):\footnote{%
We note that $G^{-1}$ can be simply inverted because the
usual geometric series truncates after three terms:
\begin{equation}
(G_C^{ - 1}  + \Sigma )^{ - 1} 
= G_C  - G_C \Sigma G_C  + G_C \Sigma G_C \Sigma G_C\,.
\end{equation} }
\begin{equation}
G = \left( {\begin{matrix}
s - {R}s^2  + T^2 u s^2
& - {T}us
&T^2us^2 - R s^2 
\\
{ - Tus}
& u
& -Tus
\\
{T^2}us^2 - Rs^2
& -Tus
& -s -R s^2 + T^2 us^2 \\
\end{matrix}} \right)\,.\eqn{Gpoles}
\end{equation}
Thus we see that there are light double poles (the $s^2$ terms),
but no higher order poles. 
A more detailed analysis shows that the residues of the double poles
vanish when $m_A=m_S$.

This completes the argument for a single valence quark.
What if there are two or more valence quarks? At the quark level,
the addition of extra valence quarks does not change any of
the elements of the block of $G$ we consider. Thus the final form of $G$, 
 \eq{Gpoles}, is unchanged, and the same double poles are present.
The argument holds separately for each valence quark,
and thus $G_{BBBB}$ has double poles of the same form
as $G_{AAAA}$, etc., although
$R$ and $s$ depend on the mass of the valence quark.
The argument can be generalized to study the structure
of $G_{AABB}$ (which has a double pole when $m_A=m_B$),
but we do not give details here.
\end{subsection}

\begin{subsection}{PQ Chiral Effective Theory}

The previous argument concerned two-point functions of quark bilinears
$\Pi_{ij}$. In the derivation of the structure of $G$ (\eq{Gstruc}), however,
the Lagrangian of PQ QCD played no role, and the only properties of $\Pi_{ij}$
that were used were the transformation properties under vector
symmetries. These properties are shared by the meson fields of PQ chiral
perturbation theory, and therefore the propagators of these fields must also
have the structure \eq{Gstruc}. Moreover, the statements about the analytic
structure were based on the interpretation of the different components of $G$
(the distinction between ``connected'' and ``disconnected'',''charged'' and
``neutral'') which in turn used the language of Feynman diagrams and the 
tracing of quark lines. As discussed in section~\ref{sec:phi0}, 
the effective theory can be formulated with $\Phi_0$, 
in which case the use of this language is still justified.

It follows that in PQ chiral perturbation theory, with $\Phi_0$ included,
if the propagators for the charged mesons have only single poles, and the self
energy function is analytic at low momentum, then there are double poles in
the neutral propagators, and no higher singularities. Since the limit
$m_0\to\infty$ is well defined for these theories, and the low energy analytic
structure is independent of it, we conclude that the above discussion also 
applies when the $\Phi_0$ field is removed from the effective theory.

Some of the assumptions and their implications are 
tested by the NLO calculation described in \cite{shshII}. There  we found
\begin{align}
& s = \frac{{Z_A }}
{{p^2  + M_{AA}^2 }},\quad u = \frac{1}
{{{{(p^2  + M_{SS}^2 )} \mathord{\left/
 {\vphantom {{(p^2  + M_{SS}^2 )} {Z_S }}} \right.
 \kern-\nulldelimiterspace} {Z_S }} + \delta _{SS}  + m_0^2 }}, \\ 
 & R = m_0^2  + \delta _{AA} ,\quad T = m_0^2  + \delta _{AS} .
\end{align}
$Z_A$ and $Z_S$ are wave function renormalization factors and $M_{AA}$ and
$M_{SS}$ meson masses, the expressions for which can be found in
\cite{shshII}. Relevant here is the fact that they are all independent of
momentum and of $m_0$, and that $M_{SS}$ and $M_{AA}$ are
light. $\delta_{SS},\ \delta_{AA},\ \delta_{AS}$ are defined through
\begin{align}
\delta _{ab}  = \tfrac{{16}}
{{f^2 }}L_7 \chi _a \chi _b  + \tfrac{1}
{{48\pi ^2 f^2 }}(p^2  - \chi _a  - \chi _b )\tfrac{1}
{2}(\chi _a  + \chi _b )\ln (\tfrac{1}
{2}(\chi _a  + \chi _b )).
\end{align}

We see that the connected contribution to the propagator for the valence quark
($s$) has only a simple light pole, while the pole of the full sea quark
$\eta'$ propagator ($u$) is heavy (at $p^2=-m_0^2+\ldots$). Also, the self
energy components ($R,\ T$) are clearly analytic in momentum. Finally, we
collect all the terms into $G$, and take the limit $m_0\to\infty$ to get
\begin{align}
G = \left( {\begin{matrix}
{s + s^2 \xi }&{ - s}&{s^2 \xi }\\
{ - s}&0&{ - s}\\
{s^2 \xi }&{ - s}&{ - s + s^2 \xi }\\
\end{matrix}} \right),
\end{align}
with
\begin{align}
\xi  =  - \delta _{AA}  + 2\delta _{AS}  - 
\delta _{SS}  - \frac{{p^2  + M_{SS}^2 }} {{Z_S }}.
\end{align}

In summary, the argument presented above serves both as a confirmation that
(given the validity of the assumptions) the pole structure of two-point
functions predicted by LO and NLO PQ chiral perturbation theory is indeed that
of PQ QCD, and an extension of that prediction to all orders in chiral
perturbation theory.

\end{subsection}

\end{section}

\begin{section}{Conclusions}
\label{sec:conclusions}

The role of $\Phi_0$ in quenched and PQ chiral perturbation theory has been
the main focus of this paper. We have shown that in order to reproduce the low
momentum behavior of two-point correlation functions of quenched QCD, $\Phi_0$
must be kept in the theory. On the other hand, in PQ QCD it does not give rise
to long range correlations, in closer analogy to the $\eta'$ in QCD, and should
not be included. This point is key in carrying out the program outlined in
\cite{shshI,ckn,shshII,shshIII} of using PQ simulations together with PQ chiral
perturbation theory to determine the unknown constants that govern the low
energy behavior of real QCD.\footnote{%
First results from this program have recently been presented~\cite{irving}.}
The central fact used in this program is that the
parameters of the chiral Lagrangian in QCD (with 2 or 3 light flavors) and in
PQ QCD (with the corresponding number of sea quarks) are the same. In the
presence of $\Phi_0$, however, the PQ chiral Lagrangian matches the unquenched
chiral Lagrangian in which the $\eta'$ field is present. The latter theory
does not have a low energy expansion (for the physical values $N=2,3$
and $N_{\rm color}=3$), and its relation to low energy hadron
phenomenology cannot be calculated perturbatively. What we have shown is that
this problem does not arise because the PQ chiral effective theory can and
should be formulated without $\Phi_0$.

In seeming contradiction to what has just been stated, there are technical
benefits from keeping $\Phi_0$ in the effective theory. In this paper we have
shown how $\Phi_0$ can be included as an auxiliary field with a mass term
$m_0^2\Phi_0^2$, and its effects can be then removed by taking $m_0$ to
infinity. This also establishes the status of previous results in PQ chiral
perturbation theory in which $\Phi_0$ was kept. By taking $m_0\to\infty$, all
effects of $\Phi_0$ are removed from these results, irrespective
of whether other $\Phi_0$ couplings were included.

The role of $\Phi_0$ is tied in with the more general theme of the foundation
and justification of PQ chiral perturbation theory. We have addressed this
issue by attempting to repeat the line of reasoning that leads to the standard
chiral Lagrangian. As a first step, we have identified the full symmetries of
PQ QCD. We then argued that the symmetry breaking pattern in this theory can
be derived from the symmetry breaking pattern of QCD. Goldstone's theorem,
with the use of the appropriately generalized Ward identities, then leads to
the conclusion that two-point 
correlation functions of operators associated with
generators of broken symmetries have low-lying poles. We discussed in some
length the construction of the effective theory for the fields that have the
same quantum numbers as these operators. This theory is guaranteed to recover
the low energy behavior of two-point functions of the chiral currents and
densities. In the absence of a Hilbert space, however, we do not know how to
show that the long range behavior of {\em general} $n$-point functions can be
attributed to the singularities of the two-point functions. This is a crucial
implicit assumption that is made when one uses 
the PQ chiral effective theory.

Lacking a general argument to justify PQ chiral effective theory,
we have focused instead on 
one of the strikingly unphysical aspects of PQ chiral perturbation theory,
the existence of light double poles in propagators of flavor-neutral mesons. 
We have demonstrated that the existence of these double poles (and
the absence of higher singularities) follows from the assumption that
the propagators of charged mesons have only simple poles.\footnote{%
This assumption can itself be derived in the massless theory
(as in section \ref{sec:lowpoles}), but not in the interesting
case of massive quarks with $m_V\ne m_S$.
}
The proof involves only the symmetries of the theory,
symmetries that are shared by the
underlying microscopic theory and the low energy effective theory.
We learn two things from this result.
First, that this unphysical feature of the effective theory is correctly
representing the properties of the underlying PQ theory.
And, second, that the pole structure seen in LO
and NLO chiral perturbation theory will hold also at higher orders---there
will only be single and double poles.

Finally, we note that an interesting consistency check of
our results can be obtained by taking the valence quark
masses to be much smaller than the sea quark masses,
though not so small that enhanced chiral logarithms\cite{shPQ}, proportional
to $m_S \ln m_V$, invalidate chiral perturbation theory.\footnote{%
This is analogous to studying the chiral $SU(2)$ theory as a limit
of chiral $SU(3)$.
We thank Larry Yaffe for suggesting this regime to us.}
In this regime the PQ theory has a ``light'' sector, with
correlators having poles at $M_{\rm light}^2\propto m_V$,
and a ``heavy'' sector with poles at $M_{\rm heavy}^2\propto m_S$.
We expect the relevant degrees of freedom in the light sector 
to be the valence quarks and ghosts alone, and thus that it should
be described by an effective {\em quenched} $SU(N_V|N_V)$ chiral Lagrangian.
In particular, this Lagrangian should contain the quenched 
$\Phi_0^Q$ field, despite the absence of the $\Phi_0$ in the underlying PQ
chiral Lagrangian. We also expect that additional terms,
such as $\alpha (\partial \Phi_0^Q)^2$, should appear in
the quenched effective Lagrangian.
This issue can be investigated analytically, 
since both valence and sea quarks are in the chiral regime.
We have checked that our expectations are indeed borne out, by
matching pole positions at one-loop obtained from the underlying
PQ theory and the effective quenched theory.
We find, for example, that a PQ theory having
$N$ degenerate sea quarks matches with a quenched theory having
$m_0^2= 2 B_0 m_S = \chi_S$ and $\alpha=1$ 
(as well as small, $m_S$ and $N$ dependent, 
shifts in $B_0$ and $f$ between the two theories).

\end{section}

\section*{Acknowledgments}
This work was partly supported by the 
U.S. Department of Energy through grant
DE-FG03-96ER40956/A006.
We are particularly grateful to Maarten Golterman and Larry Yaffe for 
detailed comments on the manuscript,
and many useful conversations.
We also acknowledge many useful conversations with Bob Singleton,
and helpful comments from Martin Savage.

\newpage

\appendix
\renewcommand\thesection{Appendix \Alph{section}}

\section{Real and Fake Symmetries}
\label{app:fakesym}

In this appendix we discuss the symmetry group of the PQ theory 
defined in \eq{Z}, 
and the resulting Ward identities. 
We show that, although the flavor symmetry group differs from
the naive expectation $SU(N_V+N|N_V)_L\otimes SU(N_V+N|N_V)_R$,
the Ward identities coincide with those derived assuming this
``fake'' symmetry to hold.
This appendix is based in part on the analogous development for
the quenched theory worked out in Ref.~\cite{GSS}.

\subsection*{Quark Sector Symmetries}

Consider first only the quark part of the action, \eq{SF},
\begin{align}
\int {(\bar \psi _L \slashed{D}\psi _L  + 
\bar \psi _R \slashed{D}\psi _R  + \bar \psi _L m\psi _R  
+ \bar \psi _R \bar m\psi _L )} 
\end{align}
In the massless limit it is invariant under transformations
of the form
\begin{align}
\psi _{L,R}  \to G_{L,R} \psi _{L,R} ,\quad \bar \psi _{L,R}  
\to \bar \psi _{L,R} G_{L,R}^{ - 1} ,\eqn{LRsym}
\end{align}
where $G_L$ and $G_R$ need only be non-singular, $G_{L,R}\in
GL(N_V+N)$. Requiring that the functional measure be invariant
reduces the symmetry to
\begin{equation}
SL(N_V+N)_L\otimes SL(N_V+N)_R \otimes U(1) \,.
\label{eq:truesymm}
\end{equation}
In the following we focus on the flavor symmetries,
and do not show the overall $U(1)$ phase symmetry.

The group~(\ref{eq:truesymm}) is larger than the symmetry group
of the Hamiltonian,
$SU(N_V+N)_L\otimes SU(N_V+N)_R$, 
because $\bar\psi \ne \psi^\dagger \gamma_0$ in the functional
integral formulation.
To understand the physical significance of the enlarged symmetry,
we consider the resulting Ward identities.
The infinitesimal transformations take the form
(choosing $G_L$ for illustration):
\begin{align}
& G_L  = \exp (\alpha T) \simeq 1 + \alpha T \\ 
 & \delta \psi _L  = \alpha T\psi _L ,\quad \delta \bar \psi _L  
=  - \alpha \bar \psi _L T,
\end{align}
where $T$ is an arbitrary traceless, hermitian $(N_V+N)^2$ matrix,
and the small parameter, $\alpha$, is complex.
The usual, unitary,  symmetry transformations
correspond to $\alpha$ being pure imaginary,
and thus have half as many parameters.
Choosing $\alpha$ to be space-time dependent,
we have (still in the massless theory)
\begin{align}
& S \to S - \int {\alpha (x)\partial _\mu  } 
j_{L\mu }^{(T)} (x),\quad j_{L\mu }^{(T)}  
= \bar \psi _L \gamma _\mu  T\psi _L , \eqn{Strans}\\ 
 & O(y) \to O(y) 
+ \alpha (y)\delta _L^{(T)} O(y),\eqn{Otrans}
\end{align}
where $\mcal{O}$ is a local operator.%
\footnote{It is convenient to use a somewhat inconsistent notation in which 
the variations of fundamental fields and 
generic operators are defined differently. 
Thus, in the case of left-handed transformations of the quark fields
\begin{align}
& \psi  \to \psi  + \delta _L \psi , \\ 
 & O \to O + \alpha \delta _L O,
\end{align}
so that, in fact, $\delta_L\mcal{O}$ is not an infinitesimal quantity.
}
Such a transformation can be seen as a 
change of variables in the functional integral, 
with unit Jacobian, and therefore leads 
(in the example of a single operator) to
\begin{align}
0 = \delta \left\langle {O(y)} \right\rangle = \int {\alpha
(x)\partial _\mu \left\langle {j_{L\mu }^{(T)} (x)O(y)}
\right\rangle dx} + \alpha (y)\left\langle {\delta _L^{(T)} O(y)}
\right\rangle .\eqn{qward}
\end{align}
Note that 
\eq{qward} contains $\alpha$ but not $\alpha^*$. 
Thus, although it appears that two equations can be obtained 
for each independent matrix $T$ (for real and imaginary $\alpha$), 
in fact only one Ward identity results:
\begin{align}
\partial _\mu \left\langle {j_{L\mu }^{(T)} (x)O(y)} 
\right\rangle = -
\delta (x - y)\left\langle {\delta _L^{(T)} O(y)} \right\rangle
\,.\eqn{Ward}
\end{align}
Thus the unitary subgroup (imaginary $\alpha$) is sufficient to
generate all the Ward identities implied by the full symmetry group.

When the masses are not zero, another term is added to \eq{qward},
corresponding to the variation in the action because of the mass terms. These
terms, however, like the operator $\mcal{O}$, do not involve complex
conjugates of the quark fields, and therefore their variation under the
transformation involves only $\alpha$. As in the massless case, this implies
that there is only one Ward identity (a modification of \eq{Ward})
corresponding to each generator $T$.

Clearly the same analysis holds for the right-handed transformations.
We conclude that the unitary subgroup,
$SU(N_V  + N)_L  \otimes SU(N_V  + N)_R $, is sufficient
to generate all the Ward identities implied by the full symmetry group.

\subsection*{Ghost Sector Symmetries}

In the PQ QCD partition function, \eq{Z}, one integrates independently
over the Grassmann fields $\psi$ and $\bar \psi$.
The integral over the commuting ghost variables, however,
converges only if it has a Gaussian structure:
\begin{align}
\int {Dx Dx^\dag \exp ( - x^\dag  Ax)}\,, 
\end{align}
where the hermitian part of $A$ 
must be positive definite. 
Since $\slashed{D}$ is anti-hermitian, 
this constraint applies to the mass term.
Thus in the ghost part of the action,
\begin{equation}
\int (\bar{\tilde \psi}_L \slashed{D} \tilde \psi_L  
+ \bar{\tilde \psi}_R \slashed{D} \tilde \psi_R  
+ \bar{\tilde \psi}_L \tilde m \tilde \psi_R  
+ \bar{\tilde \psi}_R \tilde m \tilde \psi_L )
\,,
\end{equation}
we must identify 
$\bar{\tilde\psi}_L = \tilde\psi_R^\dagger$
and $\bar{\tilde\psi}_R = \tilde\psi_L^\dagger$.
The ghost part of $S_F$ is then:
\begin{align}
\int {\tilde \psi _L^\dag \slashed{D}\tilde \psi _R + \tilde \psi _R^\dag
\slashed{D}\tilde \psi _L + \tilde \psi _L^\dag \tilde m\tilde \psi _L +
\tilde \psi _R^\dag \tilde m\tilde \psi _R }\,.
\end{align}
The symmetries of the kinetic terms alone are thus
\begin{align}
\tilde \psi _L  \to G\tilde \psi _L ,\quad \tilde \psi _R  
\to (G^{ - 1} )^\dag  \tilde \psi _R ,\eqn{ghostVtrans}
\end{align}
where $G\in GL(N_V)$. The anomaly reduces this $GL(N_V)$ symmetry group
to a product of $SL(N_V)$ and an overall phase rotation.

Ward identities are derived from infinitesimal local transformations,
$G = \exp (\alpha T) \simeq 1 + \alpha T$,
with $T$ a traceless, hermitian, $N_V\times N_V$ matrix,
and $\alpha$ complex, leading to
\begin{align}
\delta \tilde\psi_L = \alpha T \tilde\psi_L \,, \quad
\delta \tilde\psi_R = -\alpha^* T \tilde\psi_R \,,\quad
\delta \tilde\psi_L^\dag = \alpha^* \tilde \psi_L^\dag T \,,\quad
\delta \tilde\psi_R^\dag = -\alpha \tilde \psi_R^\dag T \,.
\end{align}
Local operators,
$\mcal{O}(\tilde\psi_L,\tilde\psi_L^\dag,\tilde\psi_R,\tilde\psi_R^\dag)$,
transform like
\begin{align}
\delta O(y) = \underbrace {\left( 
{\frac{{\partial O}}{{\partial \tilde\psi_L }}T\tilde\psi_L - 
\tilde\psi_R^\dag T\frac{{\partial O}}
{{\partial \tilde\psi_R^\dag }}} 
\right)}_{\equiv \delta _L O}\alpha(y)  
- \underbrace {\left( 
{\frac{{\partial O}}{{\partial \tilde\psi_R }}T\tilde\psi_R 
- \tilde\psi_L^\dag T\frac{{\partial O}}
{{\partial \tilde\psi_L^\dag}}} 
\right)}_{\equiv \delta _R O}\alpha^*(y) \,.
\end{align}
One obtains (for the case of the expectation value of a single operator)
\begin{align}
\int_x \alpha(x) 
\left\langle {\partial j_L^{(T)}(x) O(y)} \right\rangle 
- \alpha(x)^*
\left\langle {\partial j_R^{(T)}(x) O(y)} \right\rangle 
= - \alpha(y)
\left\langle {\delta_L O(y)} \right\rangle 
+ \alpha(y)^* \left\langle
{\delta _R O(y)} \right\rangle 
\,,
\end{align}
with
\begin{align}
j_{L,R\mu }^{(T)}  \equiv  
\tilde \psi_{R,L}^\dag \gamma_\mu  T \tilde \psi_{L,R} .
\end{align}
By taking $\alpha$ real and imaginary the full set of Ward 
identities are seen to be equivalent to the equations
\begin{align}
\left\langle {\partial j_{L,R}^{(T)}(x) O(y)} \right\rangle  
= - \delta(x-y) \left\langle {\delta _{L,R} O(y)} \right\rangle .
\end{align}
The generalization to $\tilde m\ne0$ is straightforward.

The resulting identities are exactly those that
one would obtain were one to pretend that $\tilde\psi$
and $\bar{\tilde\psi}$ were independent, and that the symmetry group was
$SU(N_V)_L \otimes SU(N_V)_R$. 
We note, however, that, unlike the situation in the quark sector,
the true symmetry group $SL(N_V)$ does not contain the
``fake'' symmetry group $SU(N_V)_L \otimes SU(N_V)_R$.
While they do share a common vector $SU(N_V)$ subgroup,
and have the same number of generators,
the axial transformations (left and right handed fields
rotating oppositely) differ.
We also note that the use of complex $\alpha$ is crucial for
obtaining all the independent Ward identities.

\subsection*{Graded Symmetries}

To complete the symmetries we need to consider graded transformations
which rotate Grassmann and ghost fields into each other.
Once we do this, we are necessarily considering the ghost fields to
be commuting elements of a Grassmann algebra, with
(in the notation of Ref.~\cite{deWitt}) both a ``body''
or ``base''---the usual complex scalar field---and a ``soul''---composed
of products of an even number of Grassmann fields, and thus nilpotent.
The following question thus arises: What constraint 
does the requirement of a convergent functional integral impose on
$\bar{\tilde \psi}$ and $\tilde\psi$?
We argue at the end of this Appendix that the answer is 
\begin{equation}
\bar{\tilde\psi}_L\big|_{\rm body} = \tilde\psi_R^\dagger\big|_{\rm body}
\,,\qquad
\bar{\tilde\psi}_R\big|_{\rm body} = \tilde\psi_L^\dagger\big|_{\rm body}
\,,
\label{eq:bodyrel}
\end{equation}
i.e. the relations discussed above
hold for the bodies of these quantities, but not for the souls.
In this subsection we discuss the consequences of this constraint
on the symmetries.

To do this, we return to the notation
\begin{align}
Q_{L,R}^T  = (\psi _{L,R}^T ,\tilde \psi _{L,R}^T )\,,
\quad \overline Q_{L,R}  = (\bar \psi _{L,R} ,\bar{\tilde \psi}_{L,R} )\,.
\end{align}
The most general anomaly-free flavor transformation
which leaves the kinetic term 
\begin{align}
\int {\overline Q_L \slashed{D}Q_L  + \overline Q_R \slashed{D}Q_R } \,,
\end{align}
invariant is
\begin{align}
\begin{matrix}
{Q_L }& \to &{LQ_L }&\quad &{Q_R }& \to &{RQ_R }\\
{\overline Q_L }& \to &{\overline Q_L L^{ - 1} }&
\quad &{\overline Q_R }& \to &{\overline Q_R R^{ - 1} }\\
\end{matrix}
\,.\eqn{Qtrans}
\end{align}
Here $L$ and $R$ are independent $SL(N_V+N|N_V)$ matrices,
except that they must maintain the constraint (\ref{eq:bodyrel}).
If we write
\begin{align}
L = \left( {\begin{matrix}
{L_{qq} }&{L_{qg} }\\
{L_{gq} }&{L_{gg} }\\
\end{matrix}} \right),
\end{align}
(and similarly for $L^{-1},\ R$ and $R^{-1}$) 
to denote the quark-quark block,
quark-ghost block, etc., of the different matrices,
then the constraint is
\begin{equation}
L_{gg}\big|_{\rm body} = \left(R_{gg}^{-1}\right)\big|_{\rm body}^\dag \,.
\eqn{funnygroup}
\end{equation}
In deriving this, we have used 
\begin{equation}
(L^{-1})_{gg}\big|_{\rm body} = \left(L_{gg}\big|_{\rm body} \right)^{-1}
= L_{gg}^{-1}\big|_{\rm body}
\end{equation}
and related results, which follow from the fact that only the product
of ``bodies'' contribute to the body of a product.
The matrices $L$ and $R$ satisfying the constraint \eq{funnygroup}
form a subgroup of $SL(N_V+N|N_V)_L \otimes SL(N_V+N|N_V)_R$.
The exponential parameterization of elements of this subgroup is
\begin{equation}
L = \exp(i\Phi_L) \,,\quad
R = \exp(i\Phi_R) \,,\quad
{\str\( \Phi_{L,R}\)} = 0\,,\quad
 (\Phi_L)_{gg}\big|_{\rm body}^\dag =
 (\Phi_R)_{gg}\big|_{\rm body} 
\,.
\eqn{funnyinf}
\end{equation}

To derive Ward identities we consider infinitesimal
transformations of the form in \eq{funnyinf}.
Transformations in the quark-quark and ghost-ghost blocks lead
only to the identities described in the previous subsections. 
In particular, while the full symmetry group has
axial transformations in the ghost-ghost block which were
not considered above (in which 
$(\Phi_{L,R})_{gg}$ are both pure soul) 
these do not lead to independent Ward identities.

Additional Ward identities do arise from purely graded transformations.
These are derived by considering $i\Phi_{L,R}=\alpha_{L,R} T$, 
with $T$ hermitian matrices contained entirely in the quark-ghost 
and ghost-quark blocks, and $\alpha_{L,R}$  anticommuting parameters.
Note that there are no constraints on $\alpha_{L,R}$ from
\eq{funnyinf},
and so the derivation of Ward identities follows the same steps as in
the quark sector, except that one must keep track of the anticommuting
nature of $\alpha$.
The result is a single independent identity
of the form of \eq{Ward} for each ``off-diagonal'' generator $T$.
As in the quark sector,
the identities are the same as those that follow from the
unitary subgroup in which $\Phi_{L,R}$ are constrained to be hermitian.
The extra freedom of complex parameters does not lead to additional
identities.

Combining the results from all infinitesimal transformations,
we see that all the Ward identities could have been obtained if
one had assumed the fake symmetry group 
$SU(N_V+N|N_V)_L\otimes SU(N_V+N|N_V)_R$.

\subsection*{Vector and axial transformations and currents}

The VEV of \eq{OmegatildeVEV} 
(and also the mass term with $M\propto I$)
breaks the chiral symmetry of \eq{Qtrans}
down to its vector subgroup:
\beq 
L=R\equiv V \in SL(N_V+N|N_V)\,,\qquad
V_{gg}\big|_{\rm body} = \left(V_{gg}^{-1}\right)\big|_{\rm body}^\dag \,.
\eqn{Vector} 
\eeq
The corresponding ``fake'' group, sufficient for deriving vector
Ward identities, is the subgroup 
$V \in SU(N_V+N|N_V)$, for which the
constraint in \eq{Vector} is automatically satisfied.

The axial transformations, the generators
of which are broken by the VEV, are given by
\beq
L=R^{-1}\equiv A \in SL(N_V+N|N)\,, \qquad
A_{gg}\big|_{\rm body} = \left(A_{gg}\right)\big|_{\rm body}^\dag \,.
\eqn{Axial} 
\eeq
Here the fake transformations have $A\in SU(N_V+N|N)$, 
and are not contained in the transformations of \eq{Axial} 
because the constraint is not satisfied.

By combining infinitesimal left- and right-handed transformations
in the appropriate way, we can derive vector and axial Ward identities.
They take the same form as \eq{Ward}, with $L\to V,A$, and contain
\begin{align}
j_{V,A\mu }^{(T)} (x) = \overline Q_L \gamma _\mu  TQ_L (x) 
\pm \overline Q _R \gamma _\mu  TQ_R (x)
\eqn{AxialCurrent}
\end{align}
and 
\beq
\delta_{V,A}^{(T)} O = \delta_L^{(T)} O \pm \delta_R^{(T)} O \,.
\eqn{deltaOA}
\eeq
The only subtlety here is that, when deriving Ward identities
for graded transformations, the factor of $\alpha(y)$ which is
pulled out of eqs.(\ref{eq:Strans}-\ref{eq:qward}) 
should be accompanied by the sign which results when moving $\alpha$
past $\overline Q$. This is needed to be consistent with the definition
of $j_\mu^{(T)}$, and impacts the definition of $\delta^{(T)}$.

\subsection*{Convergence considerations}

First recall that integrals over real c-numbers
lead to results of the same form as over ordinary numbers~\cite{deWitt}
\begin{equation}
\int_a^b dq\ f(q) = F(b)-F(a)\,,\quad
F' = f\,.
\eqn{gradedint}
\end{equation}
Here functions of c-numbers are defined by Taylor expansions,
\begin{equation}
f(q) = f\(q\big|_{\rm body}\)+ \sum_{n=1}^{\infty}
\frac1{n!} \left(q\big|_{\rm soul}\right)^n 
f^{(n)}\(q\big|_{\rm body}\)\,,
\end{equation}
where the sum actually truncates 
because $q\big|_{\rm soul}$ is nilpotent.
In our case we are interested in Gaussian integrals:
\begin{equation}
f(q) = \exp( -m q^2)\,,\quad
a = -\infty + a\big|_{\rm soul}\,,\quad
b = +\infty + b\big|_{\rm soul}\,,
\end{equation}
with $\mbox{Re}(m)$ positive.
Since $f^{(n)}(\pm\infty)=0$ for all $n$, it follows that
\eq{gradedint} is actually independent of the souls of $a$ and $b$.
Similarly, if we change variables by a quantity which
is pure soul, $q' = q + \delta q$, $\delta q\big|_{\rm body}= 0$,
we do not have to change the limits of integration.\footnote{%
Note that this is not true for general integrals:
one must take care when making changes of
variables involving nilpotent parts, since they can lead to
so-called anomalies~\cite{verbaarschot,Zirnbauer}.}
This is true for an arbitrary change in soul---in particular,
it does not need to be real.

Now consider a two-dimensional Gaussian integral
\begin{equation}
\int_{-\infty}^{+\infty} dq_1 dq_2 \exp\left[-m(q_1^2 + q_2^2) \right]
\,.
\eqn{twodGauss}
\end{equation}
As we have just seen, the bodies of $q_{1,2}$ are real, but their
souls can be arbitrary without changing the value of the integral.
Thus if we change to ``complex'' variables
\begin{equation}
\tilde q = q_1 + i q_2 \,,\quad
\bar{\tilde q} = q_1 - i q_2\,,\quad
d\tilde q d\bar{\tilde q} \equiv dq_1 dq_2 \,,
\end{equation}
then the integral takes on the usual complex Gaussian form
\begin{equation}
\int d\tilde q d\bar{\tilde q} \exp\( -m\bar{\tilde q} \tilde q\) \,,
\eqn{complexGauss}
\end{equation}
except that  $\bar{\tilde q}=\tilde q ^*$ holds only for the bodies,
and not for the souls (since the souls of $q_{1,2}$ are not real).
The integral itself has the value given by \eq{twodGauss},
i.e. $\pi/m$. Note that $m$ can also have an arbitrary soul---what
matters for convergence is that the real part of its body is positive.

The generalization to many complex variables is straightforward.
Let $\tilde q$ and $\bar{\tilde q}$ now represent a vector
and transposed vector, respectively.
Then the integral
\begin{equation}
\int d\tilde q d\bar{\tilde q} \exp\( -\bar{\tilde q} M \tilde q \)
\end{equation}
is convergent if $M$ (taken to have no soul for now)
is hermitian and positive,
and if $\bar{\tilde q} = \tilde q^\dag$ holds for the bodies.
To see this we diagonalize $M$: $M=U^\dag D U$,
with $D$ being diagonal and positive, and $U$ unitary.
Thus if we use the variables $\tilde q'= U \tilde q$ and
$\bar{\tilde q}' = \bar{\tilde q} U^\dag$ (which leaves the measure unchanged),
then the integral factorizes into product of integrals
of the form of \eq{complexGauss}, each of which is convergent
as long as $\bar{\tilde q}'\big|_{\rm body} = \tilde q'\big|_{\rm body}^\dag$.
This relation is maintained by the unitary transformation
back to the unprimed fields.
The argument goes through if $M$ has an arbitrary soul, since the
integrand can be expanded in powers of this soul, and each term
is convergent.
Note that the general symmetry transformation, \eq{Qtrans}, maintains the
hermiticity of the body of $M$.


\section{The diagonal generators of $U(N_V+N|N_V)$}
\label{app:defs}

Here we collect some useful results concerning the generators of
graded symmetry groups. These are represented by the hermitian
$(2 N_V+N)^2$ matrices labeled $T$ in the foregoing. 
Note that the same generators serve for both the true and fake symmetry
groups. We consider here the properties of the diagonal generators.

Let $\lambda_a$ be the $N_V+N-1$ diagonal generators of
$SU(N_V+N)$, chosen to satisfy 
$\operatorname{tr} (\lambda _a \lambda _b ) = \delta _{ab}$. 
Similarly, let $\widetilde\lambda_a$ be the $N_V-1$ diagonal generators of
$SU(N_V)$, normalized in the same fashion. 
We define $T_a,\ a=1,\ldots,N+2N_V-2$ to be the set of matrices:
\begin{align}
\begin{matrix}
\hfill {N_V  + N\text{ quark indices\{ }}\\
\hfill {N_V \text{ ghost indices\{ }}\\
\end{matrix}\left( {\begin{matrix}
{\lambda _a }&\vline & 0\\
\hline
0&\vline & 0\\
\end{matrix}} \right),\quad \left( {\begin{matrix}
0&\vline & 0\\
\hline
0&\vline & {\widetilde\lambda _a }\\
\end{matrix}} \right),
\end{align}
where we use a schematic notation.
They satisfy
\begin{align}
\operatorname{str} (T_a ) = \left\{ {\begin{matrix}
{\operatorname{tr} (\lambda _a )}\\
{-\operatorname{tr} (\widetilde\lambda _a )}\\
\end{matrix}} \right\} = 0
\end{align}
and
\begin{align}
\operatorname{str} (T_a T_b ) = \left\{ \begin{gathered}
  \operatorname{tr} (\lambda _a \lambda _b ) \\ 
   - \operatorname{tr} (\widetilde\lambda _a \widetilde\lambda _b ) \\ 
  0 \\ 
\end{gathered}  \right\} = \delta _{ab} \varepsilon _a .
\end{align}

Two more generators need to be defined to complete the 
basis for the diagonal part of $U(N_V+N|N_V)$.

\subsubsection*{Partially quenched ($N\ne 0$)}
We choose
\begin{align}
T_{2N_V  + N - 1}  = \tfrac{1}
{{2\sqrt {NN_V (N_V  + N)} }}(N\bar {I} - (2N_V  + N)I)
\end{align}
and
\begin{align}
T_{2N_V  + N}  = \tfrac{1}
{{\sqrt N }}I,
\end{align}
where
\begin{align}
\bar I \equiv \left( {\delta _{ab} \varepsilon _a } \right) 
= \left( {\begin{matrix}
I&\vline & 0\\
\hline
0&\vline & { - I}\\
\end{matrix}} \right).
\end{align}
The last element, $T_{2N_V+N}$, has non-vanishing supertrace,
and generates the anomalous $U(1)$ factor of $U(N_V+N|N_V)$.

Considering now all $2N_V+N$ generators, it is straightforward to check that
\begin{align}
\left( {\operatorname{str} (T_a T_b )} \right) \equiv \left( {g_{ab} } \right)
 = \operatorname{diag} (\underbrace {1, \ldots ,1}_{N_V  + N - 1},
\underbrace { - 1, \ldots , - 1}_{N_V  - 1}, - 1,1).
\eqn{PQstr}
\end{align}

\subsubsection*{Quenched ($N=0$)}
In this case we choose
\begin{align}
T_{2N_V  - 1}  &  = I \\ 
  T_{2N_V  }  &  = \tfrac{1}
{{2N_V }}\bar I.
\end{align}

While the identity is straceless, $\bar I$ is not, and
is taken as the generator of the anomalous $U(1)$ factor. \Eq{PQstr} becomes
\begin{align}
\left( {\operatorname{str} (T_a T_b )} \right) 
\equiv \left( {g_{ab} } \right) = \left( {\begin{matrix}
1&{}&{}&{}&{}&{}&{}&{}\\
{}& \ddots &{}&{}&{}&{}&{}&{}\\
{}&{}&1&{}&{}&{}&{}&{}\\
{}&{}&{}&{ - 1}&{}&{}&{}&{}\\
{}&{}&{}&{}& \ddots &{}&{}&{}\\
{}&{}&{}&{}&{}&{ - 1}&{}&{}\\
{}&{}&{}&{}&{}&{}&0&1\\
{}&{}&{}&{}&{}&{}&1&0\\
\end{matrix}} \right).\eqn{Qstr}
\end{align}

\section{Another argument concerning $\Phi_0$}
\label{app:alter}

In this appendix we give an alternative
argument concerning the status of $\Phi_0$ in 
quenched and PQ chiral perturbation theory.
We consider two-point functions of pseudoscalar densities,
\beq
G^{(T,T')}(x) = \langle \phi^{(T)}(x) \phi^{(T')}(0) \rangle \,,
\eeq
where $T$, $T'$ run over all the generators of $U(N_V+N|N_V)$
(and thus include the identity).
For the diagonal generators, we use the basis given in \ref{app:defs}.
We assume that all quark and ghost masses are equal,
although they do not have to vanish.\footnote{%
Note that the PQ theory differs from the unquenched theory (with $N$
flavors)
even if all masses are equal. This is because the extra fields
in the PQ theory allow one to separately determine certain Wick contractions
which always arise in certain linear combinations in the unquenched
theory.}

Consider first the PQ theory.
As long as $N\ge 2$,
some of the generators lie entirely in the sea-quark sector.
Choosing $T$ and $T'$ of this form,
we can use chiral perturbation theory for QCD-like theories to infer that
the resulting
correlator has a pseudo-Goldstone boson (PGB) pole if $T=T'$,
while flavor symmetry implies that it vanishes if $T\ne T'$.
Thus we know that
\beq
G^{(T,T')}(x) = \delta_{T,T'} G_{\rm PGB}(x) \,,\qquad
T,T' \in T_{\rm sea} \,.
\eeq
We note for future reference
that $G_{\rm PGB}(x)$ is ``connected'' in the sense that 
the only  contractions which contribute are those in which
the two bilinears are connected by quark propagators.

We can extend this result to all the straceless generators
using the graded vector symmetry (which, as argued in the
text, is not spontaneously broken), with the result
\beq
G^{(T,T')}(x) = \str(T T') G_{\rm PGB}(x) \qquad \str(T) = \str(T') = 0
\,.
\label{eq:PQPGB}
\eeq
This can be shown either directly using the symmetry, along
the lines of \ref{app:Gstructure}, or by a direct comparison of the
contributing contractions. In the latter case, the overall
factor $\str(T T')$, which can be of either sign (see \ref{app:defs}),
accounts for the signs arising from fermionic Wick contractions.
The result (\ref{eq:PQPGB}) shows that there are PGB poles in
the correlation functions for each of the generators of $SU(N_V+N|N_V)$.
This agrees with the result obtained in \ref{sec:lowpolesfromsymm}
and implies that the corresponding fields should be included in the
effective Lagrangian.

Now consider the correlation functions involving the interpolating
field corresponding to $\Phi_0$,
which are obtained by setting $T$ and/or $T'$ to the identity. 
Flavor symmetry implies that if $T=I$ and $\str(T')=0$, or vica-versa,
then the correlator vanishes.
If $T=T'=I$, then, by inspecting contractions,
one can show that the correlator is proportional to that
for the $\eta'$~\cite{BGPQ}:
\beq
G^{(I,I)}(x) = \langle \Phi_0(x) \Phi_0(0) \rangle
\propto \langle \eta'(x) \eta'(0) \rangle
\eeq
where the $\eta'$ is the flavor singlet meson {\em in the sea sector}.
Valence quark and ghost contributions completely cancel.
Now, due to the axial anomaly, we know that the $\eta'$ correlator
does not have a light pole.
Thus none of the two-point functions involving $\Phi_0$ have
PGB poles. Conversely, to describe the long distance parts of
the two-point functions in the PQ theory we do not need to
consider correlators involving $\Phi_0$.

We now contrast this analysis with that for the quenched theory.
By examining contractions, or using the graded flavor symmetries,
one can show that
\beq
G_Q^{(T,T')}(x) = \str(T T') G_{\rm conn}(x) 
+ \str(T) \str(T') G_{\rm disc}(x) 
\,,
\eqn{QPGB}
\eeq
where ``conn'' and ``disc'' refer to connected and disconnected
contractions, and $T$ and $T'$ run over the generators of $U(N_V|N_V)$.
Since there is no sea sector in the quenched theory, we cannot rely
on experience with QCD to imply that there are PGB poles in some channels.
Instead, based on numerical data, we assume that there is such a pole
in the connected correlator $G_{\rm conn}(x)$. 
Then we see from (\ref{eq:QPGB}) that, if we use the basis explained
in \ref{app:defs}, there is a PGB pole in correlators
corresponding to all the generators of $SU(N_V|N_V)$ except $T=I$.
This if because, for such generators,
$\str(T^2)\ne 0$ (so the first term in \eq{QPGB} is present) 
but $\str(T)=0$  (so the second term is absent).
However, for $T=T'=I$, both $\str(T)$ and $\str(T T')$ vanish,
and so $G_Q^{(I,I)}(x) = 0$. On the other hand,
if $T=I$ and $T'=\bar I$ (the anomalous generator), 
or vica-versa, then $\str(T T')\ne 0$,
and so
\beq
G_Q^{(I,\bar I)} \propto G_{\rm conn}(x) \,.
\eeq 
Thus this cross-correlator also has a PGB pole. 
Finally, if $T=T'=\bar I$, then
\beq
G_Q^{(\bar I,\bar I)} \propto G_{\rm disc}(x)\,,
\eeq 
about which the present analysis says nothing, 
although quenched chiral perturbation
theory predicts a double PGB pole.

The important conclusion is that, to include all channels which
have PGB poles in them one must include both
$\phi^{(\bar I)}$ and $\phi^{(I)}\propto\Phi_0$.


\section{$G^{-1}$ in the limit $m_0\to\infty$}
\label{app:Ginverse}
In this section we calculate the propagator 
$G_{(\sigma )} $
in the large $m_0$ limit, where the inverse propagator is given by
\begin{align}
G_{(\sigma )}^{ - 1}  = \left( {\begin{matrix}
A&B\\
C&{m_0^2  + d}\\
\end{matrix}} \right),
\end{align}
with $A,\ B,\ C,$ and $d$ independent of $m_0$. Note that $A$ is a square
matrix, $B$ and $C$ a column and row vectors, respectively, and $d$ a number.
Since $d$ appears only in the combination $m_0^2+d$ it can be dropped in the
large $m_0$ limit.

We write
\begin{align}
G_{(\sigma )}  = \left( {\begin{matrix}
{A'}&{B'}\\
{C'}&{d'}\\
\end{matrix}} \right).
\end{align}
To learn about the $m_0$ dependence of the blocks of $G_{(\sigma)}$ we
consider the equation
\begin{align}
G_{(\sigma )} G_{(\sigma )}^{ - 1}  = I.\eqn{GGinv}
\end{align}
One of the equations contained in \eq{GGinv} is
\begin{align}
C'A + d'C = 0,
\end{align}
which implies $d'\sim C'$, 
where the tilde refers to the scaling with $m_0$.

Another equation in \eq{GGinv} is
\begin{align}
C'B + d'm_0^2  = 1.\eqn{a'B'}
\end{align}
Since $d'$ and $C'$ scale the same, 
only the second term on the left hand 
side of \eq{a'B'} is important when $m_0\to\infty$, 
and we conclude that
\begin{align}
d',C' \sim \frac{1}
{{m_0^2 }}.
\end{align}
Similarly, from
\begin{align}
A'B + B'm_0^2  = 0
\end{align}
we get
\begin{align}
B' \sim \frac{1}
{{m_0^2 }}A',
\end{align}
which is then used in
\begin{align}
A'A + B'C = 1\quad \xrightarrow[{m_0  \to \infty }]{}\quad A'A = 1.
\end{align}
In the last equation we see that $G_{(\sigma)}^{-1}$ cannot be inverted when
$m_0\to\infty$ unless $A$ is a non-singular matrix.

Putting everything together, in the large $m_0$ limit
\begin{align}
& A' = A^{ - 1}  + \mathcal{O}\left( {\frac{1}
{{m_0^2 }}} \right) = \mathcal{O}(1), \\ 
 & B',C',d' = \mathcal{O}\left( {\frac{1}
{{m_0^2 }}} \right),
\end{align}
and therefore
\begin{align}
\mathop {\lim }\limits_{m_0  \to \infty } G_{(\sigma )}  
= \left( {\begin{matrix}
{A^{-1}}&0\\
0&0\\
\end{matrix}} \right).
\end{align}

\section{The Structure of the Propagator from Graded Symmetries}
\label{app:Gstructure}

In this appendix we derive the constraints on the structure of the pion
propagator that imply the form \eq{Gstruc}. The following symmetries are used:
\begin{itemize}
\item 
Independent phase rotations of individual flavors. These form a subgroup of
the vector transformations, \eqs{Qtrans}{Vector}, where one takes
\begin{align}
V = \operatorname{diag} (\exp \theta _A ,\exp \theta _S ,
\exp \theta _{\tilde A} ),
\end{align}
and $\theta_A$, $\theta_S$ and $\theta_{\tilde A}$ are independent. 
Under a phase rotation of only the flavor $m$,
\begin{align}
G_{ijkl}  \to G_{ijkl} \exp 
(\theta_m (\delta _{im}  - \delta _{jm}  + \delta _{km}  - \delta _{lm} )).
\eqn{U1}
\end{align}

\item 
$SU(1|1)$ transformations that involve only $A$ and $\tilde A$. These too are
vector transformations, this time with
\begin{align}
V = \left( {\begin{matrix}
a&0&b\\
0&1&0\\
c&0&d\\
\end{matrix}} \right),
\end{align}
where
\begin{align}
U = 
\left( {\begin{matrix}
a&b\\
c&d\\
\end{matrix}} \right) \in \operatorname{SU} (1|1).
\end{align}
\item $G_{ijkl}=(-)G_{klij}$. This symmetry follows from
\begin{align}
\begin{split}
{\left\langle {\Pi _{ij} \left( x \right)\Pi _{kl} \left( 0 \right)} \right\rangle }&{ = \left\langle {\Pi _{ij} \left( { - x} \right)\Pi _{kl} \left( 0 \right)} \right\rangle }\\
{}&{ = \left\langle {\Pi _{ij} \left( 0 \right)\Pi _{kl} \left( x \right)} \right\rangle }\\
{}&{ = ( - )\left\langle {\Pi _{kl} \left( x \right)\Pi _{ij} \left( 0 \right)} \right\rangle ,}\\
\end{split}\eqn{rotation}
\end{align}
where the first equation is obtained by rotation, the second by translation,
and the $(-)$ sign in the third is only needed when both $\Pi_{ij}$ and
$\Pi_{kl}$ are fermionic fields.
\end{itemize}

The invariance of $G$ under transformations of the form \eq{U1} implies that
the indices must be paired up (quark lines must be followed, corresponding to
legal contraction of quark fields). The non-vanishing elements of $G$
therefore take the form $G_{iijj}$ or $G_{ijji}$. The implications of the
$SU(1|1)$ symmetries are slightly less straightforward. We first form the
following 2-indexed objects out of the elements of $G$:
\begin{align}
\begin{split}
{\sum\limits_{j = A,\tilde A} {G_{ijjk} (x)} }& = {\left\langle {(\Pi (x)\Pi (0))_{ik} } \right\rangle }\\
{\sum\limits_{j = A,\tilde A} {\varepsilon _j G_{jjik} (x)} }& = {\left\langle {\operatorname{str} (\Pi (x))\Pi _{ik} (0)} \right\rangle }\\
{G_{SSik} (x)}& = {\left\langle {\Pi _{SS} (x)\Pi _{ik} (0)} \right\rangle }\\
{G_{iSSk} (x)}& = {\left\langle {\Pi _{iS} (x)\Pi _{Sk} (0)} \right\rangle .}\\
\end{split}\eqn{2ind}
\end{align}
The matrix notation ($\Pi$ with no indices, matrix multiplication, and the
strace symbol) refers to $2\times 2$ matrices in the $A-\tilde A$ subspace. In
a consistent manner, $i,k\in\{A,\tilde A\}$. All of the combinations above
transform similarly under $SU(1|1)$ transformations:
\begin{align}
O_{ik}  \to \sum\limits_{m,n =  A,\tilde A } {U_{im} O_{mn} U_{nk}^\dag  } ,
\end{align}
or, using the $2\times 2$ notation again,
\begin{align}
O \to UOU^\dag  .
\eqn{SU11}
\end{align}
Since $SU(1|1)$ is a symmetry, each of the combinations are invariant
under these transformations.
Direct examination shows that this implies that each $O$ 
must be proportional to the identity. From this
argument we obtain the following set of equations:
\begin{align}
& \sum\limits_{j = A,\tilde A} {G_{ijjk} }  = 
r\delta _{ik}  \Rightarrow \left\{ \begin{gathered}
  G_{AAAA}  + G_{A\tilde A\tilde AA}  = r \hfill \\
  G_{\tilde AAA\tilde A}  + G_{\tilde A\tilde A\tilde A\tilde A}  
= r \hfill \\ 
\end{gathered}  \right. \\ 
 & \sum\limits_{j = A,\tilde A} {\varepsilon _j G_{jjik} }  
= s\delta _{ik}  \Rightarrow \left\{ \begin{gathered}
  G_{AAAA}  - G_{\tilde A\tilde AAA}  = s \hfill \\
  G_{AA\tilde A\tilde A}  - G_{\tilde A\tilde A\tilde A\tilde A}  
= s \hfill \\ 
\end{gathered}  \right. \\ 
 & G_{SSik}  = t\delta _{ik}  \Rightarrow G_{SSAA}  
= G_{SS\tilde A\tilde A}  = t \\ 
 & G_{iSSk}  = v\delta _{ik}  \Rightarrow G_{ASSA}  
= G_{\tilde ASS\tilde A}  = v.
\end{align}

Finally, with the use of \eq{rotation}, these equations can be solved to
yield: 
\begin{align}
& G_{AAAA}  = r+s \\ 
 & G_{\tilde A\tilde A\tilde A\tilde A}  = r-s \\ 
 & G_{AA\tilde A\tilde A}  = G_{\tilde A\tilde AAA}  = r \\ 
 & G_{\tilde AAA\tilde A}  =  - G_{A\tilde A\tilde AA}  = s \\ 
 & G_{SSAA}  = G_{AASS}  = G_{SS\tilde A\tilde A}  
= G_{\tilde A\tilde ASS}  = t \\ 
 & G_{ASSA}  = G_{SAAS}  = G_{\tilde ASS\tilde A}  
=  - G_{S\tilde A\tilde AS}
= v.
\end{align}
The last independent element of $G$ is $G_{SSSS}$. The form shown in
\eq{Gstruc} follows (what appears there is the restriction of
$G$ to the subspace of $AA$, $SS$ and $\tilde A\tilde A$).

\newpage

\bibliography{PQ}
\bibliographystyle{h-physrev3}

\end{document}